\documentclass[a4paper, 10pt]{report}

\usepackage[cp1250]{inputenc}
\usepackage[T1]{fontenc}
\usepackage{amsfonts}
\usepackage{graphicx}
\usepackage{amsmath}

\usepackage{amssymb}

\usepackage{fancyhdr}
\usepackage{bibtopic}
\usepackage{color}
\usepackage{ulem}

\usepackage{soul}

\AtBeginDocument{}
\usepackage{etoolbox}
\patchcmd{\thebibliography}{\chapter*}{\section*}{}{}

\makeatletter
\renewcommand{\thesection}{%
  \ifnum\c@chapter<1 \@arabic\c@section
  \else \thechapter.\@arabic\c@section
  \fi
}
\makeatother

\patchcmd{\tableofcontents}{\chapter*}{\section*}{}{}

\numberwithin{equation}{section}

\begin{document}

{\LARGE \bf

\centerline{Inflation driven by scalar field and solid matter}

}

\vskip 1cm
\begin{center}
{Peter M\'esz\'aros}

\vskip 2mm {\it Department of Theoretical Physics, Comenius
University, Bratislava, Slovakia}

\vskip 4mm July 24, 2018 \vskip 1cm {\bf Abstract}
\end{center}

\vskip 2mm \noindent Solid inflation is a cosmological model where inflation is driven by fields which enter the Lagrangian in the same way as body coordinates of a solid matter enter the equation of state, spontaneously breaking spatial translational and rotational symmetry. We construct a simple generalization of this model by adding a scalar field with standard kinetic term to the action. In our model the scalar power spectrum and the tensor-to-scalar ratio do not differ from the ones predicted by the solid inflation qualitatively, if the scalar field does not dominate the solid matter. The same applies also for the size of the scalar bispectrum measured by the non-linearity parameter, although our model allows it to have different shapes. The tensor bispectra predicted by the two models do not differ from each other in the leading order of the slow-roll approximation. In the case when contribution of the solid matter to the stress-energy tensor is much smaller than the contribution from the scalar field, the tensor-to-scalar ratio and the non-linearity parameter are amplified by factors $\epsilon^{-1}$ and $\epsilon^{-2}$ respectively.

\tableofcontents

\section{Introduction}

Amongst numerous inflationary models there is a significant
subgroup of single-field ones. In the simplest of them the primordial
perturbations generated during inflation have a nearly flat spectrum and
a small level of non-Gaussianity which arises only from non-linearities
of the Einstein--Hilbert action and higher powers appearing in the potential of the scalar field
\cite{maldacena}. On the other hand, in more complicated single-field models
a significant level of non-Gaussianity is generated. For instance, models
with non-canonical kinetic term \cite{senatore} produce ''orthogonal''
non-Gaussianity described by bispectrum with positive peak at ''equilateral''
configuration of momenta ($k_1 \approx k_2 \approx k_3$) and negative
peak at ''folded'' configuration ($k_1 \approx k_2+k_3$). Examples with particular
choices of non-canonical kinetic term are $k$-inflation \cite{armendarizpicon}
and Dirac--Born-Infeld (DBI) inflation \cite{silversteintong,alishahiha}
with ''equilateral'' shape of bispectrum. Models with higher derivative
interactions may generate either ''folded'' \cite{senatore,bartolo} or
''equilateral'' type of non-Gaussianity. The latter case includes ghost
inflation \cite{arkanihamed} and models arising from effective field
theories \cite{cheung}. ''Folded'' shape of bispectrum appears in models
with non-Bunch-Davies vacuum \cite{chen,holmantolley} as well.

\vskip 2mm Multi-field models of inflation lead to ''local'' non-Gaussianity
peaking at ''squeezed'' configuration of momenta ($k_1 \approx k_2 \gg k_3$).
One of less standard examples of such models is {\it solid inflation}
\cite{gruzinov,endlich}, driven by three-component scalar field
$\phi^I$ which enter the Lagrangian in the same way as body coordinates of
solid matter enter the equation of state, so that the matter
action has to be invariant under internal translations and rotations,
\begin{eqnarray}
& & \phi^I \to M^I_{\phantom{I}J}\phi^J, \qquad \phi^I \to \phi^I
+ C^I, \nonumber\\
& & M^I_{\phantom{I}J} \in SO(3), \qquad C^I\in\mathbb{R}^3,
\qquad I,J=1,2,3, \label{eq:001}
\end{eqnarray}
where the capital indices are raised and lowered by the Euclidean metric.
The simplest possible background configuration,
\begin{eqnarray}
\phi^I=\delta^I_i x^i, \label{eq:002}
\end{eqnarray}
$x^i$ being spatial coordinates, breaks the spatial translational and rotational symmetry,
but in a flat universe it is invariant under the combined spatial-internal transformations.
As shown by Endlich et al. \cite{endlich}, in this model there appears anisotropic dependence
of the scalar bispectrum  on how the squeezed limit is approached.
Further development of the theory includes \cite{endlich2,akhshik,bartolo2,sitwell}.

\vskip 2mm Apart from the inflationary models the idea of
solid matter as one of the matter components
present in the universe was studied
in an attempt to give an alternative explanation of the
accelerated expansion of the universe, see
\cite{bucherspergel,battyemoss1,battyemoss2,battyemoss3,battyepearson}.
This can be obtained by replacing the dark energy
with a solid with negative pressure to energy density ratio
and an important example of how such solid can
be materialized are cosmic strings and domain walls \cite{battyebucher,leitemartins,kumar}.

\vskip 2mm In this paper we study a combined inflationary model including
scalar field $\varphi$ with standard kinetic term and three-component
scalar field $\phi^I$ with symmetries defined above. Similar approach
can be found in \cite{ricciardone} where the authors study a model
with special form of equation of state of the solid but non-trivial
coupling of scalar fields to gravity. In our Lagrangian
\begin{eqnarray}
\mathcal{L}=\frac{1}{2}g^{\mu\nu}\partial_\mu\varphi\partial_\nu\varphi+F(\varphi,X,Y,Z),
\label{eq:003}
\end{eqnarray}
there is no non-trivial coupling of scalar fields to gravity, but we keep the
form of the equation of state as general as possible,
omitting derivative couplings only.
Variables $X$, $Y$ and $Z$ are three independent quantities invariant
under transformations (\ref{eq:001}), for which we adopt definitions from \cite{endlich}:
\begin{eqnarray}
X=B^{II}, \quad Y=\frac{B^{IJ}B^{IJ}}{X^2}, \quad Z=\frac{B^{IJ}B^{IK}B^{JK}}{X^3},
\quad B^{IJ}=-g^{\mu\nu}\partial_\mu\phi^I\partial_\nu\phi^J,
\label{eq:004}
\end{eqnarray}
where $B^{IJ}$ is the body metric. (We have changed its sign
in order to reconcile it with
the signature of the metric tensor $(+ - - -)$, which we use throughout the paper.)
Our model represents a straightforward combination of the solid inflation and the basic single-field models.

\vskip 2mm In section 2 we study evolution of the universe in which the cosmological perturbations are absent, section 3 contains summary of the perturbation theory including the quadratic actions for the scalar and tensor perturbations, and the detailed analysis of the scalar perturbations can be found in section 4, where also the scalar spectrum is derived. Section 5 is dedicated to the scalar bispectrum and the analysis of the tensor perturbations including the tensor spectrum and bispectrum is presented in section 6. The theory explained in sections 2, 3 and 5 is based mainly on the works \cite{maldacena,endlich,weinberginin}. In the last section we discuss our main results.

\section{The unperturbed universe}

In this section we provide an analysis of our inflationary model
for the unperturbed case with flat Friedmann--Robertson--Walker--Lemaître (FRWL) metric
\begin{eqnarray}
ds^2=dt^2-a^2\delta_{ij}dx^idx^j,
\label{eq:005}
\end{eqnarray}
where $a=a(t)$ is the scale factor. The invariants $X$, $Y$ and $Z$ are then
\begin{eqnarray}
X=3a^{-2}, \quad Y=1/3, \quad Z=1/9.
\label{eq:006}
\end{eqnarray}
For two variables describing the universe, scale factor $a(t)$ and scalar
field $\varphi(t)$, we have equations of motion serving as background
equations in the perturbed theory,
\begin{eqnarray}
\dot{\varphi}^2-6M_\textrm{Pl}^2 H^2 &=& 2F, \label{eq:007} \\
\dot{\varphi}^2+2M_\textrm{Pl}^2 \dot{H} &=& 2a^{-2}F_X, \label{eq:008} \\
\ddot{\varphi}+3H\dot{\varphi} &=& F_\varphi, \label{eq:009}
\end{eqnarray}
where the subscripts stand for partial derivatives, $H=\dot{a}/a$ is the Hubble parameter
and the dot denotes the time derivative.
Of course, due to the Bianchi identity only two of these equations are independent.
We can also see that in the unperturbed case the model is described by
only two fields $X$ and $\varphi$, the former given in terms of
the scale factor by the first relation in (\ref{eq:006}).
This is a manifestation of convenience of the definition (\ref{eq:004}).
Note that the single-field model with potential dependent on
the scale factor, which uses the same variables,
differs from our model even in the absence of perturbations.

\begin{figure}[htb]
\centering
\includegraphics[scale=0.25]{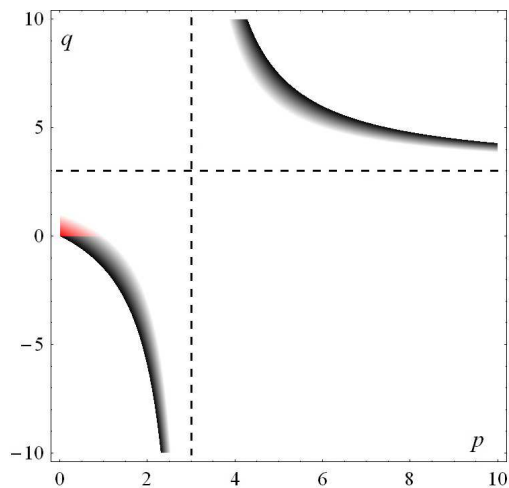}
\caption{
The slow-roll parameter is small near the hyperbolic contour given by relation (\ref{eq:011}).
The hyperbola has asymptotics at $p=3$ and $q=3$. The only region which is not
excluded by condition that Hamiltonian of scalar pertubations is bounded from below
is near the centre $p=q=0$ and is depicted by red color. This will be explained in section 3.
}
\label{fig:01}
\end{figure}

\vskip 2mm For different functions $F$ there are different solutions of the background
equations and a useful quantity measuring the deviation from the de Sitter solution
is the slow-roll parameter
\begin{eqnarray}
\epsilon=-\frac{\dot{H}}{H^2}.
\label{eq:010}
\end{eqnarray}
Using equations (\ref{eq:007}) and (\ref{eq:008}) we find
\begin{eqnarray}
\epsilon=p+q-\frac{1}{3}pq, \quad p=\frac{\dot{\varphi}^2}{2M_\textrm{Pl}^2H^2},
\quad q=X\frac{F_X}{F},
\label{eq:011}
\end{eqnarray}
where $p$ and $q$ are the slow-roll parameters of the single-field inflation
and the solid inflation respectively. In our combined model we have an additional
degree of freedom, so that the slow-roll parameter can be small also for
finite values of the parameters $p$ and $q$. The region of the parameter space
in which the slow-roll parameter is small and positive 
(so that superinflation is excluded) is depicted in fig. \ref{fig:01}.
\vskip 2mm
Relation (\ref{eq:011}) can be rewritten in terms of pressure to energy density ratios as
\begin{eqnarray}
2w-1=w_\varphi+w_s-w_\varphi w_s,
\label{eq:012}
\end{eqnarray}
where $w_\varphi=2p/3-1$ and $w_s=2q/3-1$ denote pressure to energy density ratio of the
scalar field and the solid respectively, while $w=2\epsilon/3-1$ is the overall pressure to
energy density ratio of the system consisting of these two components.
Similarly as with the dependence of $\epsilon$ on $p$ and $q$,
we can see that $w$ can be close to $-1$, in order to allow nearly de Sitter background solution,
not only if $w_\varphi\approx w_s\approx -1$ but also for a wide range of parameters $w_\varphi$ and $w_s$,
as long as the relation $w_\varphi+w_s-w_\varphi w_s\approx-3$ is satisfied.

\vskip 2mm The inflationary expansion, either exponential or power-law \cite{lucchin,liddle},
requires that the slow-roll parameter $\epsilon$
is not only small, but also has small enough time derivative. Thus, we need
another slow-roll condition, $|\eta|\ll 1$, where
\begin{eqnarray}
\eta=\frac{\dot{\epsilon}}{\epsilon H}.
\label{eq:013}
\end{eqnarray}
With the the use of the background equations and definitions of slow-roll parameters we find
\begin{eqnarray}
1+\frac{1}{2}\eta-\epsilon=\frac{1}{3\dot{\varphi}^2-2XF_X}\left(
\sqrt{\frac{6M_\textrm{Pl}^2}{\dot{\varphi}^2-2F}}
\left(3F_\varphi-XF_{X\varphi}\right)\dot{\varphi}
+2\left(X^2F_{XX}-3\dot{\varphi}^2\right)
\right).
\label{eq:014}
\end{eqnarray}

\vskip 2mm
Evolution of the unperturbed universe is determined
by initial conditions imposed on variables $a(t)$ and $\varphi(t)$ and by
function $F$, which without perturbations is effectively a function of these two variables,
$F=F(\varphi,X)$, where $X=3a^{-2}$. We do not specify this function, we only demand
restrictions on it given by slow-roll conditions.
The inflation ends when the solution $\{a(t),\varphi(t)\}$
leaves the region in $a$-$\varphi$-space in which the slow-roll conditions
on function $F$ are satisfied.

\vskip 2mm
In the single-field inflation the slow-roll parameter $\epsilon$
equals to parameter $p$ defined by (\ref{eq:010}) which
measures breaking of time reparametrization symmetry related
to time dependence of the scalar field, and the scalar field
can be used as a physical `clock' measuring the time when inflation ends.
This is different for solid inflation, since fields $\phi^I$ driving the
inflation do not depend on time, see (\ref{eq:002}).
The role of physical `clock' then must be played by metric \cite{endlich}.
In our combined model it can be played by both of them.
This leads to no contradictions as long as for given initial conditions and function $F$
the solution $\{a(t),\varphi(t)\}$ is fully specified, and therefore $\varphi$ is given
by $a$ and vice versa.

\vskip 2mm Attempting to make the analysis of the inflationary solutions more transparent,
we have replaced $X=3a^{-2}$ and $\varphi$ by $p$ and $q$, which are functions of time as well.
On the other hand, the theory under consideration is effectively described by a function
of two variables $F(X,\varphi)$. Therefore, it is possible that different functions $F(X,\varphi)$
lead to the same solution $\{X(t),\varphi(t)\}$, while it is also possible in principle that for
some functions $X$ and $\varphi$ the corresponding function $F$ does not exist,
so that such pair of functions cannot be the background solution.
The reason why the latter case cannot happen is 
summarized in the following paragraphs.

\vskip 2mm Function $F$ corresponding to a given background solution
$\{X(t),\varphi(t)\}$ can be constructed geometrically. First we define
curve $\Gamma$ in the $X$-$\varphi$-$F$-space as
\begin{eqnarray}
\Gamma: t\mapsto[X(t),\varphi(t),F(t)],
\label{eq:015}
\end{eqnarray}
where $F(t)$ can be found with the use of equation (\ref{eq:007}),
and $F$, $\varphi$ and $t$ must be replaced by dimensionless quantities
$\tilde{F}=M_\textrm{Pl}^{-4}F$,
$\tilde{\varphi}=M_\textrm{Pl}^{-1}\varphi$
and $\tilde{t}=M_\textrm{Pl}t$,
but for simplicity
we skip the tilde over them in this paragraph.
The curve $\Gamma$ is constructed to belong to a surface
given by the constraint $F=F(X,\varphi)$, so that we know the function
$F(X,\varphi)$ for $X$ and $\varphi$ being the given background solution.
For a given curve $\Gamma$ there are obviously infinitely many functions $F$,
but not all of them lead to the desired solution $\{X(t),\varphi(t)\}$,
because the gradient of this function $[F_X,F_\varphi]$
is restricted by equations (\ref{eq:008}) and (\ref{eq:009}).
As a result, the surface given by function $F$ must contain
not only curve $\Gamma$ but also an infinitesimally shifted curve
\begin{eqnarray}
\Gamma^{(\varepsilon)}: t\mapsto\left[X+\varepsilon\frac{F_X}{|\partial F|},
\varphi+\varepsilon\frac{F_\varphi}{|\partial F|},
F+\varepsilon|\partial F|\right],
\quad |\partial F|=\sqrt{F_X^2+F_\varphi^2},
\label{eq:016}
\end{eqnarray}
where all quantities in the square brackets except for
infinitesimally small $\varepsilon$ are functions of time obtained with the
use of the background equations (\ref{eq:007})-(\ref{eq:009}).

\vskip 2mm The construction of function $F$ fails if projections
of curves $\Gamma$ and $\Gamma^{(\varepsilon)}$
to the $X$-$\varphi$-plane intersect while curves themselves do not,
since at points where this happens the value of $F$ cannot be defined.
However, the Bianchi identity prevents such case to happen and the construction
of $F$ never fails.
In this way the function
$F(X,\varphi)$ is defined uniquely only for $X$ and $\varphi$ which are
infinitesimally close to the background solution.
Outside this region it can be defined in any way,
even being discontinuous.
This freedom of choice arises because our construction of function $F$
is suited to a background solution with specific initial conditions.

\vskip 2mm Letting the initial conditions be arbitrary, the question of existence
of a function $F$ generating a given system of solutions $\{X(X_0,t),\varphi(\varphi_0,t)\}$
changes considerably.
For example solutions with an exponential growth of the scale factor, $a(t)\propto \exp{(Ht)}$,
($H$ being cosntant)
and the scalar field as a linear function of time, $\varphi(t)\propto t$, can be obtained
if $p(t)$ and $q(t)$ are constant functions such that $q=3p/(p-3)$.
Function $F$ generating such solutions has to satisfy conditions
\begin{eqnarray}
F=M_\textrm{Pl}^2 H^2 (p-3),
\quad F_X=3M_\textrm{Pl}^2 H^2 p /X,
\quad F_\varphi=\pm 3\sqrt{2}M_\textrm{Pl} H^2 \sqrt{p},
\label{eq:019}
\end{eqnarray}
so that, if $p\neq 0$, function $F$ must be constant while its gradient is non-zero.
Although such function does not exist, one can find a function satisfying these conditions
for $X$ and $\varphi$ belonging to an arbitrary curve in the $X$-$\varphi$-plane,
and we can chose this curve to be the desired background solution.
The conclusion is that there exists a function $F$ which generates an exponential growth
of the scale factor and a linear dependence of the scalar field on the time only for
special initial conditions $\{a_0,\varphi_0\}$ and for any other initial conditions
the form of the background solution must be different.

\vskip 2mm A trivial example of function $F$ generating solutions of the form
$a\propto\exp{(Ht)}$ and $\varphi=\textrm{const.}$ regardless of initial conditions
is the constant function $F=-3M_\textrm{Pl}^2 H^2$, corresponding to parameters $p$ and $q$
which are both zero.
One could find a way how to construct $F(X,\varphi)$
from arbitrary $\{X(X_0,t),\varphi(\varphi_0,t)\}$,
which in contrast to our construction may not always be possible,
but the result which we have obtained
here will suffice for the purpose of the rest of our work.

\section{Quadratic actions}

For the perturbation theory
we adopt the technical approach from \cite{maldacena}
and we use the spacialy flat slicing gauge as in \cite{endlich}.
This section contains a summary of the technicalities
of this approach and the results obtained for our model within it.
The most important results to be used in the following sections
are the quadratic actions for the tensor and scalar perturbations
(\ref{eq:033}) and (\ref{eq:040}),
and useful additional relations (\ref{eq:042})-(\ref{eq:047b}).
The scalar and tensor cubic action can be found in the following sections
where bispectra are computed.

\vskip 2mm Leaving the unperturbed case, the flat FRWL metric (\ref{eq:005})
must be replaced by a more general one. In the ADM parametrisation
the general metric is
\begin{eqnarray}
ds^2=N^2dt^2-h_{ij}(dx^i+N^idt)(dx^j+N^jdt).
\label{eq:020}
\end{eqnarray}
The components of the inverse metric are
\begin{eqnarray}
g^{00}=N^{-2}, \quad g^{0i}=-N^{-2}N^i, \quad g^{ij}=-h^{ij}+N^{-2}N^iN^j,
\label{eq:021}
\end{eqnarray}
and the flat FRWL metric corresponds to $N=1$, $N^i=0$ and $h_{ij}=a^2\delta_{ij}$.
For analyzing perturbations we will use spacialy flat slicing gauge
in which the scalar and vector perturbations of the three-dimensional metric are set to zero.
The perturbed metric is then given by
\begin{eqnarray}
N=1+\delta N, \quad N^i=\xi_{,i}+N^i_T, \quad
h_{ij}=a^2\exp(\gamma_{ij}),
\label{eq:022}
\end{eqnarray}
where $N^i$ is decomposed into scalar and vector parts (longitudinal and transversal parts
in Helmholtz decomposition), $N^i_T$ satisfying $N^i_{T,i}=0$,
and $\gamma_{ij}$ is traceless and transversal,
i.e. $\gamma_{ii}=0$ and $\gamma_{ij,j}=0$.
Similarly, for perturbations of the inflationary fields we have
\begin{eqnarray}
\label{eq:023}
\varphi=\varphi_0+\delta\varphi, \quad \phi^I=\delta^I_ix^i+\pi^I,
\quad \pi^I=\rho_{,I}+\pi^I_T,
\end{eqnarray}
where $\pi^I_{T,I}=0$.

\vskip 2mm The overall action of our inflationary model consisting of the Einstein--Hilbert and matter part is 
\begin{eqnarray}
S=\int\sqrt{-g} d^4x\left[\frac{1}{2}M_\textrm{Pl}^2R
+\frac{1}{2}g^{\mu\nu}\partial_\mu\varphi\partial_\nu\varphi+F(\varphi,X,Y,Z)\right],
\label{eq:024}
\end{eqnarray}
and can be rewritten as
\begin{eqnarray}
S=\int d^4xN\sqrt{h}\left\{
\frac{1}{2}M_\textrm{Pl}^2\left[R^{(3)}+\frac{1}{N^2}\left(E^i_jE^j_i-(E^i_i)^2\right)\right]
+\frac{1}{2}\partial_\mu\varphi\partial^\mu\varphi+F
\right\},
\label{eq:025}
\end{eqnarray}
where we have followed notations of \cite{maldacena}.
$R^{(3)}$ denotes the three-dimensional scalar curvature
corresponding to the spatial metric $h_{ij}$ and
the extrisic curvature of equal-time hypersurfaces is
\begin{eqnarray}
K_{ij}=N^{-1}E_{ij}=\frac{1}{2N}\left(\dot{h}_{ij}-\nabla_iN_j-\nabla_jN_i\right),
\label{eq:026}
\end{eqnarray}
with the covariant derivative with respect to the spatial metric denoted by $\nabla$.
By varying this action with respect to $N^i$ and $\delta N$ we obtain the momentum
and hamiltonian constraints,
\begin{eqnarray}
M_\textrm{Pl}^2\nabla_j\left[N^{-1}(E^j_i-\delta^j_iE^k_k)\right]
+N\partial_{N^i}\left(\frac{1}{2}\partial_\mu\varphi\partial^\mu\varphi+F\right)&=&0, \\
\label{eq:027}
\frac{1}{2}M_\textrm{Pl}^2\left[R^{(3)}-\frac{1}{N^2}\left(E^i_jE^j_i-(E^i_i)^2\right)\right]+
\partial_{\delta N}\left(\frac{1}{2}N\partial_\mu\varphi\partial^\mu\varphi+NF\right)&=&0.
\label{eq:028}
\end{eqnarray}
Considering $\delta N$, $\xi$ and $N^i_T$ in the form of the plane waves with the wavenumber $k$,
these constraints are satisfied up to the first order of the perturbation theory for
\begin{eqnarray}
\label{eq:029}
\delta N & = & \frac{(M_\textrm{Pl}^2Hk^2\dot{\varphi}_0-F_XF_\varphi)\delta\varphi
+\dot{\varphi}_0F_X\dot{\delta\varphi}
+2a^{-2}F_X^2k^2\rho-2M_\textrm{Pl}^2HF_Xk^2\dot{\rho}}
{2FF_X+2M_\textrm{Pl}^4H^2k^2}, \\
\label{eq:030}
\xi & = & \frac{-(F\dot{\varphi}_0+M_\textrm{Pl}^2HF_\varphi)\delta\varphi
+M_\textrm{Pl}^2H\dot{\varphi}_0\dot{\delta\varphi}
+2M_\textrm{Pl}^2Ha^{-2}F_Xk^2\rho+2FF_X\dot{\rho}}
{2FF_X+2M_\textrm{Pl}^4H^2k^2}, \\
\label{eq:031}
N^i_T & = & \frac{4F_X\delta^i_I\dot{\pi}^I_T}{4F_X-M_\textrm{Pl}^2k^2}.
\end{eqnarray}
Knowing the higher order corrections is not necessary unless the fourth order terms
in the action are needed, since the second order terms of $N$ and $N^i$
multiply the first order constraint equations and their third order terms multiply
the zeroth order constraints \cite{maldacena}.

\vskip 2mm Expanding the action (\ref{eq:025}) up to the second order and using
(\ref{eq:029})-(\ref{eq:031}) we can find the quadratic action which determines the evolution
of $\delta\varphi$, $\rho$, $\pi^I_T$ and $\gamma_{ij}$ in first
order of the perturbation theory. We can also use integration by parts together
with relations $\pi^I_{T,I}=0$, $N^i_{T,i}=0$, $\gamma_{ii}=0$ and $\gamma_{ij,j}=0$
appearing in definitions of these perturbations. In this way we obtain the quadratic action
decomposed into three parts
\begin{eqnarray}
S^{(2)}=S^{(2)}_S+S^{(2)}_V+S^{(2)}_T,
\label{eq:0032}
\end{eqnarray}
where the scalar, vector and the tensor parts are denoted
by $S^{(2)}_S$, $S^{(2)}_V$ and $S^{(2)}_T$ respectively.

\vskip 2mm The tensor quadratic action can be written in the form
\begin{eqnarray}
S^{(2)}_T=\frac{1}{4}M_\textrm{Pl}^2\int d^4xa^3\left(
\frac{1}{2}\dot{\gamma}_{ij}\dot{\gamma}_{ij}
-\frac{1}{2}a^{-2}\gamma_{ij,k}\gamma_{ij,k}+2\tilde{h}c^2_T\gamma_{ij}\gamma_{ij}
\right),
\label{eq:033}
\end{eqnarray}
where $\tilde{h}=\dot{H}+\dot{\varphi}_0^2/(2M_\textrm{Pl}^2)=H^2(p-\epsilon)$
and $c_T$ denotes the transverse sound speed,
\begin{eqnarray}
c^2_T=1+\frac{2}{3}\frac{F_Y+F_Z}{XF_X}.
\label{eq:034}
\end{eqnarray}
The vector quadratic action is
\begin{eqnarray}
S^{(2)}_V=M_\textrm{Pl}^2\int\frac{d^3kdt}{(2\pi)^3}a^3\left(
\frac{1}{4}\frac{k^2}{1-k^2/(4a^2\tilde{h})}\dot{\pi}^I_T\dot{\pi}^I_T
+\tilde{h}c^2_Tk^2\pi^I_T\pi^I_T
\right).
\label{eq:035}
\end{eqnarray}
The quadratic terms in actions are written in the simplified form.
Terms such $\textrm{Re}\left\{\psi_{\mathbf{k}}\chi^*_{\mathbf{k}}\right\}$
we denote as $\psi\chi$, and we use this notation also in the rest of this section.

\vskip 2mm The scalar quadratic action can be written as a sum of three parts as
\begin{eqnarray}\label{eq:036}
S^{(2)}_S&=&M_\textrm{Pl}^2\int\frac{d^3kdt}{(2\pi)^3}a^3\left[
\frac{1}{3}\frac{\tilde{e}k^4}{1-\tilde{e}k^2/(3a^2\tilde{h})}
\left(\dot{\rho}-\frac{\tilde{h}}{H}\rho\right)^2
+\tilde{h}c^2_Lk^4\rho^2
\right]+\\
&+&S^{(2)}_{\delta\varphi}+S^{(2)}_{\delta\varphi\textrm{-}\rho},
\nonumber
\end{eqnarray}
where $\tilde{e}=1-\dot{\varphi}_0^2/(2F)=3/(3-p)$, $c_L$
is the longitudinal sound speed,
\begin{eqnarray}
c^2_L=1+\frac{2}{3}\frac{XF_{XX}}{F_X}+\frac{8}{9}\frac{F_Y+F_Z}{XF_X},
\label{eq:037}
\end{eqnarray}
and $S^{(2)}_{\delta\varphi}$ denotes action quadratic in $\delta\varphi$,
while $S^{(2)}_{\delta\varphi\textrm{-}\rho}$ is action consisting of
$\delta\varphi$-$\rho$-type terms. The first part of action (\ref{eq:036})
is written in the explicit form for the sake of being easily compared to
(6.4) in \cite{endlich} as the special form
of our action with parameter $p$ set to zero.
The same applies to the tensor and vector parts.

\vskip 2mm The full scalar quadratic action written
with coefficients expressed in terms of the slow-roll
parameter $\epsilon$ and the parameter $p$ is
\begin{eqnarray}\label{eq:038}
S^{(2)}_S &=& \int\frac{d^3kdt}{(2\pi)^3}a^3\biggl\{
M_\textrm{Pl}^2\tilde{Q}k^4\left(\dot{\rho}+HQ\rho\right)^2
-M_\textrm{Pl}^2H^2Qc^2_Lk^4\rho^2+\\
&+& \frac{1}{2}\left(1+p\tilde{Q}\right)\dot{\delta\varphi}^2
-\sqrt{p}\tilde{p}S_{+}\delta\varphi\dot{\delta\varphi}+\nonumber\\
&+& \frac{1}{2}\left(F_{\varphi\varphi}-\frac{k^2}{a^2}+\left[\left(\frac{1}{2}\eta_p-Q\right)k^2
+SS_{+}\right]H\sqrt{p}\tilde{p}\right)\delta\varphi^2+\nonumber\\
&+& \left(-2F_{X\varphi}\pm\sqrt{2}M_\textrm{Pl}\sqrt{p}\tilde{Q}S_{+}\right)
\frac{k^2}{a^2}\delta\varphi\rho\pm\nonumber\\
&\pm& \sqrt{2}M_\textrm{Pl}\sqrt{p}\tilde{Q}k^2\left[
H\left(\frac{1}{2}\eta_p-Q\right)\delta\varphi\dot{\rho}-
HQ\dot{\delta\varphi}\rho-
\dot{\delta\varphi}\dot{\rho}
\right]
\biggr\},\nonumber
\end{eqnarray}
where
\begin{eqnarray}
\label{eq:039}
& & Q=\epsilon-p, \quad \tilde{Q}=\frac{a^2H^2Q}{M+k^2},
\quad \tilde{p}=\frac{H\sqrt{p}}{M+k^2}, \quad M=a^2H^2(3-p)Q, \\
& & \eta_p=\frac{\dot{p}}{pH}, \quad S=3-\epsilon+\frac{1}{2}\eta_p,
\quad S_{+}=a^2H^2QS+k^2. \nonumber
\end{eqnarray}
In order to have a proper sigh of this action, in the sense that the
corresponding Hamiltonian is bounded from below, both $\tilde{Q}$
and $Q$ must be possitive. Demanding possitivity of them for
all values of the wavenumber $k$, we find the restriction $Q>0$,
i.e. $0<p<\epsilon$. This considerably norrows the parameter space of the theory,
so that only the red region in diagram in fig. \ref{fig:01} is allowed.

\vskip 2mm
Note that the $\pm$ sign in front of the last term in the action originates from
expressing $\dot{\varphi}_0$ in terms of slow-roll parameters,
$\dot{\varphi}_0=\sqrt{2}M_{\textrm{Pl}}H(\pm\sqrt{p})$.
The plus sign ($+\sqrt{p}$) corresponds to case when $\varphi_0$ grows during inflation
while the case when $\varphi_0$ decreases corresponds to minus sign ($-\sqrt{p}$).
Keeping both of these cases the sign $\pm$ will appear throughout the rest of the paper.
\vskip 2mm

Unfortunately, equations governing evolution of perturbations
obtained by variation of this action are coupled.
Due to the effect of gravity, this occurs even if
$F_{X\varphi}$ is set to zero.
In order to quantize scalar perturbations properly,
$\delta\varphi$ and $\rho$ then must be replaced by their linear combinations
$\delta\tilde{\varphi}$ and $\tilde{\rho}$ such that
the part of the action
$S^{(2)}_{\delta\tilde{\varphi}\textrm{-}\tilde{\rho}}$ describing coupling vanishes.
Finding the transformation relation
$(\delta\varphi,\rho)\mapsto(\delta\tilde{\varphi},\tilde{\rho})$,
or finding the solution for $\delta\varphi$ and $\rho$ directly,
is a matter of solving a complicated system of differential equations.
The problem can be simplified in the special case if $p$
is small, at most of the same order as $\epsilon$,
and $F_{X\varphi}$
is of a higher order,
when the action written up to the next-to-leading order
in the slow-roll approximation reduces to
\begin{eqnarray}
\label{eq:040}
S^{(2)}_S &=& \int\frac{d^3kdt}{(2\pi)^3}a^3\biggl[
M_\textrm{Pl}^2a^2H^2(k^2-3a^2H^2Q)Q\dot{\rho}^2+
\\
&+& 2M_\textrm{Pl}^2a^2H^3k^2Q^2\dot{\rho}\rho-
M_\textrm{Pl}^2H^2c_L^2k^4Q\rho^2+
\nonumber\\
&+& \frac{1}{2}\dot{\delta\varphi}^2+\frac{1}{2}\left(
F_{\varphi\varphi}-\frac{k^2}{a^2}+3H^2p \right)\delta\varphi^2
-Hp\delta\varphi\dot{\delta\varphi} \pm
\nonumber\\
&\pm& \sqrt{2}M_\textrm{Pl}a^2H^2\sqrt{p}Q \left(\frac{k^2}{a^2}\delta\varphi\rho-
\dot{\delta\varphi}\dot{\rho}\right)
-2F_{X\varphi}\frac{k^2}{a^2}\delta\varphi\rho
\biggr].\nonumber
\end{eqnarray}
As a consequence of the restriction $p<\epsilon$ following from analysis
of signs of terms in the action (\ref{eq:038}), the parameter $\eta_p$ defined by
the fifth relation in (\ref{eq:039}) can be expressed as
\begin{eqnarray}
\label{eq:dod1}
\eta_p=\frac{\epsilon}{p}\eta+\frac{1}{3}\frac{F_{\varphi\varphi}}{H^2},
\end{eqnarray}
where only the leading order terms of the slow-roll approximation have been kept.
Moreover, $F_{X\varphi}$ being much smaller than the slow-roll parameter
yields $F_{\varphi\varphi}\sim\epsilon$, and therefore parameter $\eta_p$
is small as well.

\vskip 2mm For the inflationary expansion of the universe, the smallness of parameter $p$
requires also smallness of parameter $q$, because up to the first order in the slow-roll
parameter, relation (\ref{eq:011}) is simplified to $\epsilon=p+q$, and considering
smallness of $\eta_p=\dot{p}/(pH)$, $\dot{q}/(qH)$ must be small as well.
For this reason not only $\epsilon$ and $\eta$, but also $p$ and $\eta_p$
may be called slow-roll parameters.
Consequently, higher derivatives of the function $F$
with respect to $X$ and $\varphi$ cannot be arbitrary either.
By differentiating parameters $p$ and $q$ with respect to time and using
the background equations (\ref{eq:007})-(\ref{eq:009}) we find that
if $p$ is not much greater than $\epsilon$
the partial derivatives of $F$
are constrained by slow-roll parameters,
\begin{eqnarray}
\label{eq:042}
XF_X,X^2F_{XX},X^3F_{XXX},\sqrt{p}F_{\varphi}\sim\epsilon,\qquad
X\sqrt{p}F_{X\varphi},X^2\sqrt{p}F_{XX\varphi},...
\sim\epsilon^2.
\end{eqnarray}

\vskip 2mm Up to the leading order in slow-roll parameters
the sound speeds (\ref{eq:034}) and (\ref{eq:037}) can be rewritten in the form
\begin{eqnarray}
\label{eq:043}
c^2_T=1+\frac{2}{3}\frac{F_Y+F_Z}{XF_X}, \quad
c^2_L=\frac{1}{3}+\frac{1}{9}\frac{p}{\epsilon-p}\frac{F_{\varphi\varphi}}{H^2}
+\frac{8}{9}\frac{F_Y+F_Z}{XF_X},
\end{eqnarray}
and neglecting the first order terms of the slow-roll approximation we obtain constraints
\begin{eqnarray}
\label{eq:047}
\frac{4}{3}c^2_T-c^2_L=1,\quad -\frac{3}{8}\leq\frac{F_Y+F_Z}{XF_X}\leq0,
\end{eqnarray}
so that the transverse sound speed must be greater than $\sqrt{3}/2$
and the longitudinal one must be smaller than $\sqrt{1/3}$, unless $\epsilon-p\sim\epsilon^2$.
The assumption of real sound speeds was also taken into account,
since if they were imaginary
the undesired exponential growth of the perturbations would occur.
We can also see that $\epsilon-p$ cannot be
much smaller than $\epsilon^2$, more precisely without neglecting 
the term with $H^{-2}F_{\varphi\varphi}$ in (\ref{eq:043}) we obtain
\begin{eqnarray}
\label{eq:047b}
-3\leq\frac{p}{\epsilon-p}\frac{F_{\varphi\varphi}}{H^2}\leq18,\quad -\frac{3}{2}\leq\frac{F_Y+F_Z}{XF_X}\leq0.
\end{eqnarray}

\section{Scalar perturbations}

By varying the quadratic action (\ref{eq:040}) we obtain
equations for scalar perturbations in the form of plane waves
with wavenumber $k$,
\begin{eqnarray}
\label{eq:045}
& &\ddot{\rho}_k+\left(5-2\epsilon+\eta_Q-6\frac{a^2H^2}{k^2}Q\right)H\dot{\rho}_k
+\left[\left(5+3c_L^2\right)Q+\frac{k^2c_L^2}{a^2H^2}\right]H^2\rho_k
=\nonumber\\
& &=\pm\frac{\sqrt{p}}{\sqrt{2}M_\textrm{Pl}k^2}\left[\ddot{\delta\varphi}_k+5H\dot{\delta\varphi}_k
+\frac{k^2}{a^2}\left(1\pm\frac{\sqrt{2}M_\textrm{Pl}F_{X\varphi}}{\sqrt{p}F_X}\right)\delta\varphi_k\right],
\end{eqnarray}
\begin{eqnarray}
\label{eq:046}
& &\ddot{\delta\varphi}_k+3H\dot{\delta\varphi}_k-
\left(F_{\varphi\varphi}-\frac{k^2}{a^2}+6H^2p\right)
\delta\varphi_k=\nonumber\\
& &=\pm \sqrt{2}M_\textrm{Pl}a^2H^2\sqrt{p}Q\left[\ddot{\rho}_k+5H\dot{\rho}_k+\frac{k^2}{a^2}\left(1\pm\frac{\sqrt{2}M_\textrm{Pl}F_{X\varphi}}{\sqrt{p}F_X}\right)\rho_k \right],
\end{eqnarray}
where
\begin{eqnarray}\label{eq:046_1}
\eta_Q=\frac{\dot{Q}}{QH}=\frac{\epsilon\eta-p\eta_p}{\epsilon-p}.
\end{eqnarray}
One can find the Fourier mode functions of $\delta\varphi$
and $\rho$ either by solving these
equations numerically or employing some approximative methods such as
the uniform approximation, see \cite{habib1,habib2}, but we restrict ourselves
to the case when a simple form of analytical solutions can be found.
This requires not only assumptions we have imposed so far, but also two additional ones.
The first assumption is that parameter $\eta_Q$ must be at most of the same order as slow-roll parameters,
and the second assumption concerns parameter $\eta_L$ defined as
\begin{eqnarray}
\label{eq:048}
\quad \eta_L=\frac{\dot{c}_L}{c_LH},
\end{eqnarray}
which must be small as well. Smallness of $\eta_Q$ may be violated if $\epsilon-p\sim\epsilon^2$
and in such case we have to demand $\epsilon\eta-p\eta_p\sim\epsilon^3$. Note that analytical solution
of equations for scalar perturbations can be easily found also for arbitrary value of $\eta_Q$ as
long as its time dependence is mild, but this choice results in large value of the scalar spectral tilt,
see relation (\ref{eq:064_1}) in the end of this section, which is refuted by observations.

\vskip 2mm  The right-hand side of equation (\ref{eq:046}) can be neglected if
\begin{eqnarray}
\label{eq:fico}
F_{X\varphi}=\pm\frac{\sqrt{p}F_X}{\sqrt{2}M_\textrm{Pl}}\left(c_L^2-1\right),
\end{eqnarray}
since in this special case the combination of terms in brackets represents the equation of motion
for the scalar perturbation $\rho$ (\ref{eq:045})
in the leading order of the slow-roll approximation.
Therefore, the equation of motion for the scalar field perturbation $\delta\varphi$
is decoupled and can be easily solved.
Equation for $\rho$ can be decoupled by replacing $\rho$ by the new variable $\mathcal{S}$ defined by
\begin{eqnarray}
\label{eq:kalinak}
\mathcal{S}_k=\rho_k\mp\frac{\sqrt{p}}{\sqrt{2}M_\textrm{Pl}k^2}\delta\varphi_k,
\end{eqnarray}
however, as we will see, to solve the equation for $\mathcal{S}$
is a bit more tricky than to solve the equation for $\delta\varphi$.

\vskip 2mm In order to solve equations of motion for perturbations it is useful
to introduce the conformal time $\tau$ defined in the standard way as
$\tau=\int a^{-1}dt$, $\tau\in(-\infty,0)$.
By replacing the cosmological time by it
and considering assumptions imposed above including special form of $F_{X\varphi}$ given by (\ref{eq:fico}), we find
\begin{eqnarray}
\label{eq:049}
& & \mathcal{S}_k^{\prime\prime}-\left(4+2\epsilon_c+\eta_{Q,c}-\frac{6Q_c}{k^2\tau^2}\right)\frac{1}{\tau}\mathcal{S}_k^\prime
+\left(\frac{5+3c_{L,c}^2}{\tau^2}Q_c+c_L^2(\tau)k^2\right)\mathcal{S}_k=0,\\
\label{eq:050}
& & \delta\varphi_k^{\prime\prime}-2\frac{1+\epsilon_c}{\tau}\delta\varphi_k^\prime+\left(
k^2-\frac{6p_c+H_c^{-2}F_{\varphi\varphi,c}}{\tau^2} \right) \delta\varphi_k = 0,
\end{eqnarray}
where the prime denotes the differentiation with respect to the conformal time,
$\mathcal{H}=a'/a$,
and the subscript $c$ stands for quantities evaluated at the reference
time $\tau_c$ when the longest mode of observational relevance today
with the wavenumber $k_\textrm{min}\sim H_\textrm{today}$ ($a_\textrm{today}\equiv1$)
exits the horizon, i.e.,
\begin{eqnarray}
\label{eq:051}
\left|\frac{k_\textrm{min}}{H_ca_c}\right|\sim\left|H_\textrm{today}\tau_c\right|=1.
\end{eqnarray}
Using this convention we also obtain relations
\begin{eqnarray}
\label{eq:052}
& & a=a_c\left(\frac{\tau}{\tau_c}\right)^{-1-\epsilon_c},\quad 
H=\frac{-1-\epsilon_c}{a_c\tau_c}\left(\frac{\tau}{\tau_c}\right)^{\epsilon_c}
=H_c\left(\frac{\tau}{\tau_c}\right)^{\epsilon_c},\\
& & \epsilon=\epsilon_c\left(\frac{\tau}{\tau_c}\right)^{-\eta_c},\quad 
p=p_c\left(\frac{\tau}{\tau_c}\right)^{-\eta_{p,c}},\quad 
Q=Q_c\left(\frac{\tau}{\tau_c}\right)^{-\eta_{Q,c}},\quad
c_L=c_{L,c}\left(\frac{\tau}{\tau_c}\right)^{-\eta_{L,c}}.\nonumber
\end{eqnarray}

\vskip 2mm The Fourier modes of $\delta\varphi$ and $\rho$ can be quantized
in the standard way as
\begin{eqnarray}
\label{eq:053}
\rho_{\mathbf{k}} &=& \rho^{(\textrm{cl})}_{\mathbf{k}}
a_{\mathbf{k}}+\rho^{(\textrm{cl})*}_{-\mathbf{k}} a_{-\mathbf{k}}^\dag, \\
\label{eq:054}
\delta\varphi_{\mathbf{k}} &=& \delta\varphi^{(\textrm{cl})}_{\mathbf{k}}
b_{\mathbf{k}}+\delta\varphi^{(\textrm{cl})*}_{-\mathbf{k}} b_{-\mathbf{k}}^\dag,
\end{eqnarray}
where the classical solutions obeying equations of motion
are denoted by the superscript $(\textrm{cl})$,
and the creation and annihilation operators obey commutation relations
\begin{eqnarray}
\label{eq:055}
\left[a_{k_1},a_{k_2}^\dag\right]=\left[b_{k_1},b_{k_2}^\dag\right]
=(2\pi)^3\delta^{(3)}(\mathbf{k}_1-\mathbf{k}_2).
\end{eqnarray}
Normalization of the classical solutions is determined by the equal time
commutation relations for $\delta\varphi$ and $\rho$ and their conjugate momenta
\begin{eqnarray}
\label{eq:056}
\left[\rho(\mathbf{x_1},t),\pi_\rho(\mathbf{x_2},t)\right]=
\left[\delta\varphi(\mathbf{x_1},t),\pi_{\delta\varphi}(\mathbf{x_2},t)\right]=
i\delta^{(3)}(\mathbf{x}_1-\mathbf{x}_2),
\end{eqnarray}
and it can be obtained by matching the canonically normalized fields
\begin{eqnarray}
\label{eq:057}
\delta\varphi^{(\textrm{can})}=a\delta\varphi^{(\textrm{cl})},\quad
\rho^{(\textrm{can})}=\sqrt{2}M_\textrm{Pl}Ha^2\sqrt{Q}k\rho^{(\textrm{cl})},
\end{eqnarray}
to the mode functions of the free wave function of the
Minkowski space vacuum,
$\frac{1}{\sqrt{2k}}e^{-ik\tau}$ or $\frac{1}{\sqrt{2c_Lk}}e^{-ic_Lk\tau}$
in the limit of very early time, $\tau\to-\infty$, when the modes are deep
inside the horizon, $k\ll Ha$, and the curvature of spacetime does not
affect their evolution.

\vskip 2mm The correctly normalized classical solutions
of equations (\ref{eq:049}) and (\ref{eq:050}) are
\begin{eqnarray}
\label{eq:058}
\rho_k^{(\textrm{cl})}&=&-i\frac{\sqrt{\pi}}{2\sqrt{2}}
\frac{H_c}{M_\textrm{Pl}\sqrt{Q_c}k}
(-\tau)^{\frac{5}{2}}H^{(1)}_{\frac{5}{2}}
(-c_{L,c}k\tau), \\
\label{eq:059}
\delta\varphi_k^{(\textrm{cl})}&=&
-\frac{\sqrt{\pi}}{2}H_c
(-\tau)^{\frac{3}{2}}
H^{(1)}_{\frac{3}{2}}(-k\tau),
\end{eqnarray}
where $H^{(1)}_\nu$ denote Hankel functions of the first kind,
and all parameters in equations of motion which are of the same order as slow-roll
parameters or smaller have been omitted.
This result is valid even without the restriction on $F_{X\varphi}$ (\ref{eq:fico}) taken into account and it is sufficient for calculation of the scalar bispectrum
in the leading order of the slow-roll approximation discussed in
the next section,
but the omitted parameters are needed to determine
the deviation of the scalar power
spectrum from the flat one.
Unfortunately, when these parameters are taken into account, equation (\ref{eq:049})
cannot be solved immediately because of the term proportional to $\tau^{-2}$
in the coefficient in front of $\mathcal{S}_k$.
The extra term can be removed by performing one more transformation
of dependent variable, mimicking the transformation used in \cite{endlich}.
The appropriate variable appears to be a scalar quantitiy $\mathcal{U}$
defined by the solid matter velocity $u^{(s)i}$ as
\begin{eqnarray}
\label{eq:059_0}
\mathcal{U}=H\delta u^{(s)}=a^2H(\dot{\rho}-\xi),
\end{eqnarray}
where the $\delta u$ is the scalar part of the solid matter velocity,
$u^{(s)}_i=\delta u^{(s)}_{,i}+u^{(s)T}_i$, $u^{(s)T}_{i,i}=0$,
and the term $-\xi$ in the brackets originates from lowering
the index with use the perturbed metric.
By inserting (\ref{eq:030}) into (\ref{eq:059_0})
and keeping only the relevant terms
in the slow-roll approximation we obtain
\begin{eqnarray}\label{eq:059_2}
\mathcal{U}_k=a^2H\left[\left(1-3\frac{a^2H^2}{k^2}Q\right)\dot{\mathcal{S}}_k
+HQ\mathcal{S}_k\right].
\end{eqnarray}
The quantity $\mathcal{U}$ is related to the quantity $\mathcal{R}$
defined with the use of the notation from \cite{weinberg}
(see equation (5.4.22) there) as
\begin{eqnarray}\label{eq:059_1}
\mathcal{R}_k=\frac{A_k}{2}+H\delta u_k,
\end{eqnarray}
where the signature $(- + + +)$ is used and $\delta u$ is the scalar part of velocity
of the system consisting of solid matter and scalar field.
In order to express the right-hand side of this definition
in the terms of scalar perturbations
present in our model we need the $(0-i)$ components of stress-energy
tensor up to the first order of the perturbation theory.
By inserting the resulting velocity potential into
the defintion of $\mathcal{R}$ and returning
to the signature of the metric tensor which we use, we find
\begin{eqnarray}
\label{eq:059_2_0}
\mathcal{R}=-\frac{Q}{\epsilon}\mathcal{U}\mp\frac{\sqrt{p}}{\sqrt{2}M_\textrm{Pl}H\epsilon}\delta\varphi.
\end{eqnarray}
In the case when the scalar field $\varphi$ is not present in the universe,
we simply have $\mathcal{R}=-\mathcal{U}$.

\vskip 2mm Using equation (\ref{eq:059_2}) together with (\ref{eq:045}), (\ref{eq:046}) and (\ref{eq:fico}),
the scalar quadratic action (\ref{eq:040}) can be rewritten
into a more convenient form
\begin{eqnarray}
\label{eq:059_3}
S^{(2)}_S &=& \int\frac{d^3kdt}{(2\pi)^3}a^3\biggl\{
\frac{M_\textrm{Pl}^2}{c_L^2}Q\biggl[
-\dot{\mathcal{U}}^2-2H(3-\epsilon-Q+\eta_Q)\dot{\mathcal{U}}\mathcal{U}+
\\
&+& \left(\frac{k^2}{a^2}c_L^2-(9-6\epsilon-6Q-3c_L^2Q+6\eta_Q)H^2\right)\mathcal{U}^2
\biggr]+
\nonumber\\
&+& \frac{1}{2}\dot{\delta\varphi}^2+\frac{1}{2}\left(
F_{\varphi\varphi}-\frac{k^2}{a^2}+3H^2p \right)\delta\varphi^2
-Hp\delta\varphi\dot{\delta\varphi}
\biggr\}.\nonumber
\end{eqnarray}
The sign of the kinetic term of $\mathcal{U}$ in the action is the opposite as for
$\rho$, because in the gauge which we use, $\rho$ measures the position
of the solid matter elements while $\mathcal{U}$ measures their velocity.
As a simple example, such a change of the sign appears also in the action of the
one-dimensional harmonic oscillator $S=\int dt (\dot{x}^2-\omega^2x^2)/2$,
which rewritten in terms of the velocity $v=\dot{x}$ takes the form
$S=\int dt (-\omega^{-2}\dot{v}^2+v^2)/2$.

\vskip 2mm Equation of motion for $\mathcal{U}$ obtained by varying the action (\ref{eq:059_3}) reads
\begin{eqnarray}\label{eq:059_4}
\mathcal{U}^{\prime\prime}-\frac{2+2\epsilon_c+\eta_{Q,c}-2\eta_{L,c}}{\tau}\mathcal{U}^\prime
+\left(k^2c_L^2(\tau)
+3\frac{(1+c_{L,c}^2)Q_c-2\eta_{L,c}}{\tau^2}\right)\mathcal{U}=0,
\end{eqnarray}
where the longitudinal sound speed as a conformal time dependent function
is given by the last relation in (\ref{eq:052}).
By matching the general form of the solution of this equation for the canonically
normalized field
\begin{eqnarray}\label{eq:059_5}
\mathcal{U}^{(\textrm{can})}=i\sqrt{2}\frac{M_\textrm{Pl}}{c_L}a\sqrt{Q}\mathcal{U},
\end{eqnarray}
to free wave mode function of the Minkowski space vacuum,
and applying the same procedure to scalar field perturbation $\delta\varphi$
with all small parameters taken into account up to the first order of
the slow-roll approximation, we find
\begin{eqnarray}
\label{eq:059_6}
\mathcal{U}^{(\textrm{cl})}_k&=&i\frac{\sqrt{\pi}}{2\sqrt{2}}\frac{H_cc_{L,c}}{M_\textrm{Pl}\sqrt{Q_c}}
(-\tau_c)^{-\epsilon^{(\mathcal{U})}_c}\left(1+\frac{1}{2}\eta_{L,c}-\epsilon_c\right)e^{i\frac{\pi}{2}p^{(\mathcal{U})}_c}\cdot
\\
& &\cdot(-\tau)^{\frac{3}{2}+\epsilon^{(\mathcal{U})}_c}
H^{(1)}_{\frac{3}{2}+p^{(\mathcal{U})}_c}(-c_L(\tau)(1+\eta_{L,c})k\tau),
\nonumber\\
\label{eq:059_7}
\delta\varphi^{(\textrm{cl})}_k&=&-\frac{\sqrt{\pi}}{2}H_c(-\tau_c)^{-\epsilon_c}(1-\epsilon_c)
e^{i\frac{\pi}{2}\epsilon^{(\delta\varphi)}_c}
(-\tau)^{\frac{3}{2}+\epsilon_c}
H^{(1)}_{\frac{3}{2}+\epsilon^{(\delta\varphi)}_c}(-k\tau),
\end{eqnarray}
where
\begin{eqnarray}
\label{eq:059_8}
\epsilon^{(\mathcal{U})}&=&\epsilon+\frac{1}{2}\eta_Q-\eta_L, \\
\label{eq:059_9}
p^{(\mathcal{U})}&=&p-c_L^2Q+\frac{1}{2}\eta_Q+\frac{5}{2}\eta_L, \\
\label{eq:059_10}
\epsilon^{(\delta\varphi)}&=&\epsilon+2p+\frac{1}{3}\frac{F_{\varphi\varphi}}{H^2}.
\end{eqnarray}
Note that consistency of commutation relations (\ref{eq:055}) with
commutation relations (\ref{eq:056}) requires that the classical modes
satisfy the relation $ff^{\prime *}-c.c.=ia^{-2}\propto (-\tau)^{-1-\epsilon_c}$.
For modes of the form we have found above the conequence of this condition
is that prefactors in front of Hankel function in (\ref{eq:059_6}) and (\ref{eq:059_7})
must be proportional to $(-\tau)^{3/2+\epsilon_c}$. Unfortunately this is true
only for the scalar field perturbation $\delta\varphi$, while for perurbation $\mathcal{U}$
the quantization is valid only in the limit $k\tau\to -\infty$.
On the other hand, the spectral tilt which is calculated below do not depend on the power
of $-\tau$ in the prefactor in front of the Hankel function.
This power affects only the mild time dependence of the size of
the power spectrum. Therefore, we find the method of normalization of modes we have used
sufficient for the purpose of finding results presented in this paper.

\vskip 2mm Our goal is to compute the correlation functions
of a scalar quantity $\zeta$ that parameterizes the curvature perturbations, defined as
\begin{eqnarray}
\label{eq:060}
\zeta_k=\frac{A_k}{2}-H\frac{\delta\rho_k}{\dot{\bar{\rho}}},
\end{eqnarray}
where the notation follows \cite{weinberg} again.
Expressed in term of $\delta\varphi$ and $\mathcal{U}$,
the scalar perturbation $\zeta$ in the leading order in slow-roll parameters is
\begin{eqnarray}
\label{eq:061}
\zeta=\frac{\pm\sqrt{p}}{\sqrt{2}M_\textrm{Pl}\epsilon}
\left(\frac{\dot{\delta\varphi}}{3H}-\delta\varphi\right)
+\frac{Q}{c_L^2\epsilon}\left(\frac{\dot{\mathcal{U}}}{3H}+\mathcal{U}\right),
\end{eqnarray}
and the corresponding two-point function
in the late time limit is
\begin{eqnarray}
\label{eq:062}
& &\left<0\right|\zeta_{k_1}\zeta_{k_2}\left|0\right>
=(2\pi)^3\delta^{(3)}(\mathbf{k}_1+\mathbf{k}_2)\frac{H^2_c}{4M_\textrm{Pl}^2\epsilon_c^2}
k_1^{-3}\left(\frac{\tau}{\tau_c}\right)^{2\epsilon_c+2\eta_c}\cdot\\
& &\cdot\left[
p_c\left(\frac{\tau}{\tau_c}\right)^{-\eta_{p,c}}
(-k_1\tau)^{-2\epsilon^{(\delta\varphi)}_c}
+\frac{Q_c}{c_{L,c}^5}\left(\frac{\tau}{\tau_c}\right)^{-\eta_{Q,c}+5\eta_{L,c}}
(-c_{L,c}k_1\tau)^{-2p^{(\mathcal{U})}_c}
\right].\nonumber
\end{eqnarray}
The scalar power spectrum $\mathcal{P}_\zeta(k)$ defined by
\begin{eqnarray}
\label{eq:063}
\left<0\right|\zeta_{k_1}\zeta_{k_2}\left|0\right>
=\frac{\mathcal{P}_\zeta(k_1)}{2k_1^3}
(2\pi)^5\delta^{(3)}(\mathbf{k}_1+\mathbf{k}_2),
\end{eqnarray}
is usually approximated by power-law function, $\mathcal{P}_\zeta(k)\propto k^{n_S-1}$,
where $n_S$ is the scalar spectral index, being close to one for a nearly flat spectrum.
The spectral tilt up to the leading order of the slow-roll approximation can be computed as
\begin{eqnarray}
\label{eq:064}
n_S-1=\frac{d \ln \mathcal{P}_\zeta}{d\ln k}=
-2\frac{c_{L,e}^5\sigma p_e\epsilon^{(\delta\varphi)}_c+
\left(\epsilon_e-p_e\right)p^{(\mathcal{U})}_c}
{\epsilon_e+\left(c_{L,e}^5\sigma-1\right)p_e},
\end{eqnarray}
where the subscript $e$ stands for quantities evaluated in the time when the inflation ends,
$\tau_e\approx 0^{-}$, and $\sigma$ denotes
\begin{eqnarray}
\label{eq:bugar}
\sigma=\left(\frac{\tau_e}{\tau_c}\right)^{2\left(p^{(\mathcal{U})}_c-\epsilon^{(\delta\varphi)}_c\right)}
=e^{2N_\textrm{min}\left(\epsilon^{(\delta\varphi)}_c-p^{(\mathcal{U})}_c\right)},
\quad \left\{
\begin{array}{cc}
\sigma \sim 10 & \epsilon^{(\delta\varphi)}_c>p^{(\mathcal{U})}_c \\
\sigma \sim 1 & \epsilon^{(\delta\varphi)}_c\approx p^{(\mathcal{U})}_c \\
\sigma \sim 1/10 & \epsilon^{(\delta\varphi)}_c<p^{(\mathcal{U})}_c
\end{array}
\right.
,
\end{eqnarray}
where $N_\textrm{min}$ is the minimal number of e-folds ($N_\textrm{min}\sim 60$), and
$\left(k_\textrm{max}/k_\textrm{min}\right)^{2\left(p^{(\mathcal{U})}_c-\epsilon^{(\delta\varphi)}_c\right)}$,
$k_\textrm{max}\sim 3000k_\textrm{min}$ being the maximal wavenumber
corresponding to the highest observed multipole moment of the cosmic microwave background,
and $c_{L,c}^{-2p_c^{(\mathcal{U})}}$ were replaced by one.
(For example $3000^{0.01}\dot{=}1.08$ and $0.1^{0.01}\dot{=}0.98$.)
The fifth power of the longitudinal sound speed appearing in relation (\ref{eq:064})
cannot be larger than $3^{-5/2}\dot{=}0.064$,
since the maximal value for the longitudinal sound speed allowing
inflationary expansion of the universe is $1/\sqrt{3}$.
and therefore,
if $\sigma$ is not greater than of order unity,
the dominant contribution to the spectral tilt is
\begin{eqnarray}
\label{eq:064_1}
n_S-1\approx -2p^{(\mathcal{U})}_c=2c_{L,c}^2\epsilon_c-2(1+c_{L,c}^2)p_c-\eta_{Q,c}-5\eta_{L,c},
\end{eqnarray}
where the second power of the longitudinal sound speed with maximal allowed value $1/3$
(unless $\epsilon-p\sim\epsilon^2$) has been kept,
whereas for $\sigma\gg 1$ we have
\begin{eqnarray}
\label{eq:064_12}
n_S-1\approx -2\epsilon^{(\delta\varphi)}_c=-2\epsilon_c-4p_c-\frac{2}{3}\frac{F_{\varphi\varphi,c}}{H^2_c}.
\end{eqnarray}

\vskip 2mm Our inflationary model contains two special cases.
The first one is the most simple single-field inflation
which can be obtained by taking the limit such that $p=\epsilon$,
when the scalar spectral tilt (\ref{eq:064}) reduces to
\begin{eqnarray}
\label{eq:064_2}
n_S-1\to-2\left.\epsilon_c^{(\delta\varphi)}\right|_{p_c=\epsilon_c}
=-6\epsilon_c-\frac{2}{3}\left.\frac{F_{\varphi\varphi,c}}{H_c^2}\right|_{p_c=\epsilon_c}
=-2\epsilon_c-\eta_c,
\end{eqnarray}
see also relation (54) in \cite{lyth}.
The second special case is the solid inflation model in which
the scalar field $\delta\varphi$ is not present and parameter
$p$ must be set to zero.
The corresponding spectral tilt is
\begin{eqnarray}
\label{eq:064_3}
n_S-1\to-2\left.p^{(\mathcal{U})}_c\right|_{p_c=0}=2c_{L,c}^2\epsilon_c-\eta_c-5\eta_{L,c},
\end{eqnarray}
the same which can be found in \cite{endlich}.

\section{Scalar bispectrum}

In the linear order of the perturbation
theory Gaussianity is preserved. Therefore, in order to
compute bispectrum which encodes the non-Gaussianity,
cubic terms in the action are needed.
These terms include higher partial derivatives
of the function $F$, and those appearing in (\ref{eq:042})
that are suppressed by the slow-roll parameter, may be neglected.
Moreover, from (\ref{eq:047}) follows
\begin{eqnarray}
\label{eq:073}
0 \leq F_Y+F_Z \leq -\frac{3}{8}X F_X =
\frac{9}{8}M_\textrm{Pl}^2H^2\left(\epsilon-p\right),
\end{eqnarray}
so that $F_Y+F_Z$ can be neglected as well.
However, it must be small, its time derivative may not be,
because the restriction (\ref{eq:073}) is just an inequalitity,
and therefore there are no restrictions on
$F_{XY}+F_{XZ}$ and $F_{Y\varphi}+F_{Z\varphi}$.
On the other hand, small functions with not small derivative
usually do not occur in physical problems,
so that it is reasonable to restrict ourselves
to the special case in which
\begin{eqnarray}
\label{eq:74}
F_Y+F_Z=\frac{9}{8}AM_\textrm{Pl}^2H^2\left(\epsilon-p\right),
\end{eqnarray}
where $A$ is a constant, or a slowly varying function, of order unity, $A\sim 1$.
By differentiating this equation we obtain the restriction
\begin{eqnarray}
\label{eq:075}
X\left(F_{XY}+F_{XZ}\right)\mp\frac{M_\textrm{Pl}}{\sqrt{2}}
\sqrt{p}\left(F_{Y\varphi}+F_{Z\varphi}\right)\sim\epsilon^2,
\end{eqnarray}
which is satisfied if we put
\begin{eqnarray}
\label{eq:076}
\tilde{F}=X\left(F_{XY}+F_{XZ}\right)=\pm\frac{M_\textrm{Pl}}{\sqrt{2}}
\sqrt{p}\left(F_{Y\varphi}+F_{Z\varphi}\right),
\end{eqnarray}
and allow $\tilde{F}$ to be of arbitrary order in the slow-roll parameters.
When $\epsilon-p$ is much smaller than $\epsilon$, restriction (\ref{eq:047})
is no longer valid and the second relation in (\ref{eq:047b}) must be used instead.
This obviously does not change the point of this paragraph.

\vskip 2mm
In the previous section we had to impose several restrictions on parameters
of the theory in order to be able to solve equations for scalar perturbations
analyticaly. Consequently the results for the scalar two-point function
and the corresponding spectral tilt are valid only in special case.
Fortunately in this section we will be able to calculate the the scalar
tree-point function and the corresponding bispectrum in more general case.
In addition to condition (\ref{eq:74}) and smallness of $p$ following from
analysis of signs of terms in scalar quadratic action \label{eq:038}
here we have to demand only two restrictions, smallness $\eta_Q$ defined in (\ref{eq:046_1})
and $\eta_L$ in (\ref{eq:048}).
As a reminder, the properly normalized classical mode
of the scalar perturbation $\rho$ (\ref{eq:058})
is of the order $\rho\sim(\epsilon-p)^{-1/2}$.
Now we have everything needed to keep track of orders in the
slow-roll approximation when collecting cubic terms of the action.

\vskip 2mm
By expanding the action (\ref{eq:025}) up to the
third order in scalar perturbations and keeping only
the leading order terms in the slow-roll approximation we find
\begin{eqnarray}
\label{eq:077}
S^{(3)}_S&=&\int d^4xa^3\Bigg[
-\frac{8}{81}\left(\frac{2}{3}F_Y+\tilde{F}\right)\left(\rho_{,ii}\right)^3+
\\
&+&\frac{8}{27}\left(F_Y+\tilde{F}\right)\rho_{,ii}\rho_{,jk}\rho_{,jk}
-\frac{8}{27}F_Y\rho_{,ij}\rho_{,ik}\rho_{,jk}\pm\nonumber
\\
&\pm&\frac{4\sqrt{2}}{9M_\textrm{Pl}}\frac{\tilde{F}}{\sqrt{p}}
\left(\rho_{,ij}\rho_{,ij}
-\frac{1}{3}\left(\rho_{,ii}\right)^2\right)\delta\varphi
\Bigg].\nonumber
\end{eqnarray}
This action determines the interaction Hamiltonian
responsible for the non-Gaussianity of scalar perturbations.
The scalar bispectrum is given by the three-point function
of the scalar $\zeta$, which can be computed with the use
of the in-in formalism \cite{weinberginin} as
\begin{eqnarray}
\label{eq:078}
\left<\zeta_{\mathbf{k}_1}(\tau)\zeta_{\mathbf{k}_2}(\tau)\zeta_{\mathbf{k}_3}(\tau)\right>
=-i\int\limits_{-\infty}^\tau a(\tau')d\tau'\left<0\right|\left[
\zeta_{\mathbf{k}_1}(\tau)\zeta_{\mathbf{k}_2}(\tau)\zeta_{\mathbf{k}_3}(\tau),
H_\textrm{int}(\tau')
\right]\left|0\right>,
\end{eqnarray}
where only the first order term with a single integration and a simple commutator is considered.
By inserting (\ref{eq:061}), (\ref{eq:077}) and classical modes (\ref{eq:058}) and (\ref{eq:059})
into this formula, and using the commutation relations (\ref{eq:055}),
we find the late time three-point function in the leading order of the slow-roll approximation,
\begin{eqnarray}
\label{eq:079}
\left<\zeta_{\mathbf{k}_1}(0)\zeta_{\mathbf{k}_2}(0)\zeta_{\mathbf{k}_3}(0)\right>
&=&\frac{H_c^2(2\pi)^3\delta^{(3)}(\mathbf{k}_1+\mathbf{k}_2+\mathbf{k}_3)}
{2M_\textrm{Pl}^6c^{12}_{L,c}\epsilon_c^3\left(k_1k_2k_3\right)^3}\bigg[
\tilde{Q}(\mathbf{k}_1,\mathbf{k}_2,\mathbf{k}_3)\Lambda_{k_1,k_2,k_3}+\\
&+&c_{L,c}^2\tilde{F}\left(
Q^{(2,3)}(\mathbf{k}_1,\mathbf{k}_2,\mathbf{k}_3)\Omega_{k_1,c_{L,c}k_2,c_{L,c}k_3}+2\textrm{ permutations}
\right)\nonumber\bigg],
\end{eqnarray}
where
\begin{eqnarray}
\label{eq:080}
\tilde{Q}(\mathbf{k}_1,\mathbf{k}_2,\mathbf{k}_3)&=&
\left(F_Y+\tilde{F}\right)\frac{k_1^2\left(\mathbf{k}_2\cdot\mathbf{k}_3\right)^2
+2\textrm{ permutations}}{\left(k_1k_2k_3\right)^2}-
\\
&-&3F_Y
\frac{\left(\mathbf{k}_1\cdot\mathbf{k}_2\right)\left(\mathbf{k}_1\cdot\mathbf{k}_3\right)
\left(\mathbf{k}_2\cdot\mathbf{k}_3\right)}{\left(k_1k_2k_3\right)^2}
-\frac{2}{3}F_Y-\tilde{F},\nonumber
\end{eqnarray}
\begin{eqnarray}
\label{eq:081}
Q^{(A,B)}(\mathbf{k}_1,\mathbf{k}_2,\mathbf{k}_3)=
\frac{1}{2}\frac{\left(\mathbf{k}_A\cdot\mathbf{k}_B\right)^2}
{\left(k_Ak_B\right)^2}-\frac{1}{6}, \quad A,B=1,2,3,
\end{eqnarray}
and $\Lambda_{k_1,k_2,k_3}$, $\Omega_{k_1,c_{L,c}k_2,c_{L,c}k_3}$,...
are given by the integrals
\begin{eqnarray}
\label{eq:082}
\Lambda_{k_1,k_2,k_3}&=&\textrm{Re}\Bigg\{
i\int\limits_0^\infty
\left(1-ik_1z-\frac{1}{3}k_1^2z^2\right)
\left(1-ik_2z-\frac{1}{3}k_2^2z^2\right)\\
& &\phantom{medzera}
\left(1-ik_3z-\frac{1}{3}k_3^2z^2\right)
e^{i(k_1+k_2+k_3)z}z^{-4}dz
\Bigg\},\nonumber
\end{eqnarray}
\begin{eqnarray}
\label{eq:083}
\Omega_{\mathcal{A},b,c}&=&\textrm{Re}\Bigg\{
\int\limits_0^\infty
\left(i+\mathcal{A}z\right)
\left(1-ibz-\frac{1}{3}b^2z^2\right)\\
& &\phantom{medzera}
\left(1-icz-\frac{1}{3}c^2z^2\right)
e^{i(\mathcal{A}+b+c)z}z^{-4}dz
\Bigg\},\nonumber
\end{eqnarray}
where the first index of $\Omega$ is denoted by capital caligraphical letter
in contrast to the second and third index denoted by small letters, because
the first index is the only one corresponds to wavenumber which is not multiplied
by the longitudinal sound speed.
These integrals obviously do not converge.
The divergence due to unbounded upper limit
of integration interval
at $z=\infty$ can be avoided by tilting the
integration contour, $z\to(1+i\varepsilon)z$,
with $\varepsilon\to0^+$.
This also provides projection on the right vacuum.
The divergence of the integral (\ref{eq:082}) due to the lower limit of
integration interval at $z=0$
is consumed by evaluating the real part of the integral,
however in the integral (\ref{eq:083}) a logarithmic divergence remains.
By calculating integrals
(\ref{eq:082}) and  (\ref{eq:083}) in this way, we obtain
\begin{eqnarray}
\label{eq:084}
\Lambda_{k_1,k_2,k_3}&=&
-\frac{1}{27\left(\sum_ik_1\right)^3}\Bigg[
3\sum_ik_i^6+9\sum_{i\neq j}k_i^5k_j+
12\sum_{i\neq j}k_i^4k_j^2+6\sum_{i\neq j}k_i^3k_j^3+
\nonumber\\
&+&18\left(\prod_ik_i\right)\sum_{i\neq j}k_i^2k_j+
18\left(\prod_ik_i\right)\sum_ik_i^3+
20\left(\prod_ik_i\right)^2
\Bigg],
\end{eqnarray}
\begin{eqnarray}
\label{eq:085}
\Omega_{\mathcal{A},b,c}&=&
\frac{1}{3}\mathcal{A}^3\left[\gamma_\textrm{EM}+\ln\left(-\tau_e\left(\mathcal{A}+b+c\right)\right)
+\mathcal{O}\left(\tau_e\left(\mathcal{A}+b+c\right)\right)
\right]-
\\
&-&\frac{1}{9\left(\mathcal{A}+b+c\right)^2}\Big[
b^5+2b^4c+2b^3c^2+2b^2c^3+2bc^4+c^5+\nonumber\\
&+&2\mathcal{A}\left(b^4+b^3c+b^2c^2+bc^3+c^4\right)+
2\mathcal{A}^2\left(2b^3+3b^2c+3bc^2+2c^3\right)+\nonumber\\
&+&\mathcal{A}^3\left(10b^2+17bc+10c^2\right)
+11\mathcal{A}^4\left(b+c\right)+4\mathcal{A}^5
\Big],\nonumber
\end{eqnarray}
where $\gamma_\textrm{EM}$ is the Euler--Mascheroni constant.
The second integral has been computed with integration limits
$(-\tau_e,\infty)$, $\tau_e$ being the time when the inflation ends,
which can be expressed as $\tau_e=-H_\textrm{today}^{-1}e^{-N_\textrm{min}}$.
The integral is then dominated by
\begin{eqnarray}
\label{eq:085_1}
\Omega_{\mathcal{A},b,c}=
-\frac{1}{3}\mathcal{N}_\zeta\mathcal{A}^3,
\end{eqnarray}
where $\mathcal{N}_\zeta$ is a number of the order of number of e-folds,
and we use this relation instead of (\ref{eq:085}) in what follows.
We also neglect the part of $\tilde{Q}(\mathbf{k}_1,\mathbf{k}_2,\mathbf{k}_3)$
defined by (\ref{eq:080}) which is proportional to $\tilde{F}$,
since its contribution is much smaller than the contribution from (\ref{eq:085_1}).

\begin{figure}[htb]
\centering
\includegraphics[scale=0.15]{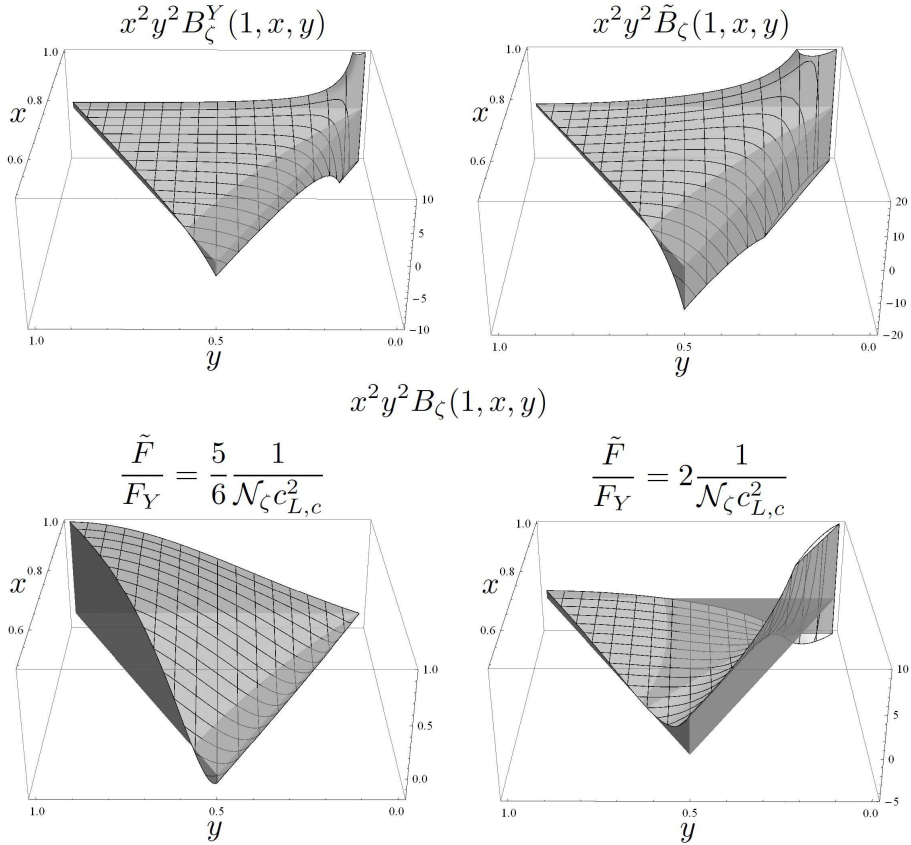}
\caption{
Shapes of the scalar bispectrum.
Flat triangles represent the zero plane.
}
\label{fig:03}
\end{figure}

\vskip 2mm As a result of computations above,
the scalar bispectrum $B_\zeta(k_1,k_2,k_3)$, defined by relation
\begin{eqnarray}
\label{eq:086}
\left<\zeta_{\mathbf{k}_1}\zeta_{\mathbf{k}_2}\zeta_{\mathbf{k}_3}\right>
=(2\pi)^3\delta^{(3)}(\mathbf{k}_1+\mathbf{k}_2+\mathbf{k}_3)
B_\zeta(k_1,k_2,k_3),
\end{eqnarray}
consists of two parts,
\begin{eqnarray}
\label{eq:087}
B_\zeta(k_1,k_2,k_3)=F_YB^Y_\zeta(k_1,k_2,k_3)
+\mathcal{N}_\zeta c_{L,c}^2\tilde{F}\tilde{B}_\zeta(k_1,k_2,k_3),
\end{eqnarray}
parametrized by three independent parameters of the theory,
$F_Y$, $\tilde{F}$ and $c_{L,c}$.
Due to the delta-function on the right-hand side of (\ref{eq:086}),
three wavenumbers $k_1$, $k_2$ and $k_3$
can be identified with the sides of a triangle,
and all information about bispectrum is encoded in a function of two
variables which characterize the shape of the triangle.
Following conventions of \cite{babich},
we define $x=k_2/k_1$ and $y=k_3/k_1$
and describe the bispectrum by the function
$x^2y^2B_\zeta(1,x,y)$
defined in region $1-x\leq y\leq x$, $1/2\leq x\leq 1$, $0\leq y\leq1$.
Shapes of the functions $x^2y^2B^Y_\zeta(1,x,y)$ and
$x^2y^2\tilde{B}_\zeta(1,x,y)$ are depicted
in the first two panels of fig. \ref{fig:03}.
All functions in the figure are normalized to have value $1$
in the equilateral limit, $x=y=1$.

\vskip 2mm Function $B^Y_\zeta(k_1,k_2,k_3)$ has the same shape as bispectrum
derived by Endlich et al. in the model where the inflation
is driven by the solid only \cite{endlich}.
It peaks in the squeezed limit, $x=1$, $y=0$, with anisotropic
dependence on how the limit is approached.
The second part of the bispectrum $\tilde{B}_\zeta(k_1,k_2,k_3)$ follows from
the presence of the scalar field in our combined model
and it has similar shape as the first one.

\vskip 2mm It is not unexpected that
our model with the additional degree of freedom
allows for a wider range of different shapes
of the bispectrum.
The overall bispectrum peaks
in the squeezed limit,
unless $\tilde{F}/F_Y=(5/6)\mathcal{N}_\zeta^{-1}c_{L,c}^{-2}$,
when it peaks in the equilateral limit instead.
This case is depicted in the third panel of fig. {\ref{fig:03}}.
An example of the overall bispectrum for
$\tilde{F}/F_Y>(5/6)\mathcal{N}_\zeta^{-1}c_{L,c}^{-2}$,
when the relative sign of the bispectrum in the squeezed limit
and the bispectrum in the equilateral limits flips,
is depicted in the fourth panel.

\vskip 2mm Apart from the shape of the bispectrum,
we are interested also in its size.
It is given by the non-linearity parameter $f_\textrm{NL}$
defined for the Newtonian potential $\Phi$, which is proportional
to the scalar $\zeta$ in the long-wavelength limit,
$\Phi=3\zeta/5$.
Following the definition (4) in \cite{creminelli}, we can use the formula
\begin{eqnarray}
\label{eq:090}
f_\textrm{NL}=\frac{5}{72\pi^4}\frac{k^6B_\zeta(k,k,k)}{\mathcal{P}_\zeta^2(k)},
\end{eqnarray}
and by inserting (\ref{eq:062}) and (\ref{eq:079}) into it, we find
\begin{eqnarray}
\label{eq:091}
f_\textrm{NL}&=&\frac{\epsilon_c}{\left[\epsilon_c+\left(c_{L,c}^5-1\right)p_c\right]^2}
\left(
\frac{19415}{13122}\frac{1}{c_{L,c}^2}\frac{F_Y}{F}-
\frac{5}{18}\mathcal{N}_\zeta\frac{\tilde{F}}{F}
\right).
\end{eqnarray}
We can see that if $\epsilon-p\sim\epsilon\sim p$,
the non-linearity parameter is of the order
$f_\textrm{NL}\sim (F_Y/F)c_L^{-2}\epsilon^{-1}$,
the same as for the solid inflation without
the scalar field,
or $f_\textrm{NL}\sim\mathcal{N}_\zeta(\tilde{F}/F)\epsilon^{-1}$.
Supposing that $c_L^5\sim\epsilon$ we have $f_\textrm{NL}\sim (F_Y/F)c_L^{-2}\epsilon^{-3}$
or $f_\textrm{NL}\sim\mathcal{N}_\zeta(\tilde{F}/F)\epsilon^{-3}$ if $\epsilon-p$ is of the order $\epsilon^2$.
The overall form of the non-linearity parameter is more complicated
than in the solid inflation, since our model
features more parameters of the theory.

\vskip 2mm The condition $\epsilon-p\lesssim\epsilon^2$ leading to an amplification of
the non-linearity parameter can be rewritten as $q\ll p$, which means that the contribution
of the solid matter to the overall stress-energy tensor is negligible in comparison
to the contribution of the scalar field. This also means that coefficients in term is quadratic action corresponding to the solid matter are negligible in comparison with coefficients of terms corresponding to the scalar field. Since coefficients of the cubic action do not depend on $\epsilon-p$, the interaction coefficient is effectively enhanced, resulting in larger non-Gaussianity. Due to smallness of coefficients in solid matter kinetic term in the quadratic action, the scalar perturbation $\rho$ is normalized as $\rho\sim(\epsilon-p)^{-1/2}$ and the non-Gaussianity is enhanced even more. This normalization also prefers interaction terms with $\rho^3$ and $\rho^2\delta\varphi$ over terms with $\rho\delta\varphi^2$ and $\delta\varphi^3$, which we have omitted in our calculations.

\vskip 2mm So far, we have two constraints on our model given by observations.
The first one concerning the scalar spectrum is that
the spectral index must have the value $n_S=0.968$ \cite{planck2},
and the second one is $f_\textrm{NL}$ to be not much larger than $10$ \cite{planck}.
In our model there are three independent
parameters of the theory which are not necessarily suppressed by the slow-roll parameters,
$F_Y$, $\tilde{F}$ and $c_L$, and in principle, the observational constraints
can be satisfied.
The way how to obtain more restrictions on our model is
to study tensor perturbations, although they are
beyond the reach of current observations.
It is only known that the tensor-to-scalar ratio
cannot be larger than of the order of $0.1$ \cite{planck2,bicep}.
However, in order to make the analysis of the model in consideration complete,
in the next section
we compute the tensor spectrum and bispectrum.

\section{Tensor perturbations}

Because the same technicalities which have been used
for the analysis of the scalar perturbations are applicable
also for the tensor perturbations, in this section
the results are summarized more succinctly
than in the previous ones.
The only results in this section which differentiate
our model from the solid inflation model are the tensor spectral
tilt (\ref{eq:099}) and the tensor-to-scalar ratio (\ref{eq:100}),
which now contain the additional slow-roll parameter $p$.
The tensor bispectrum is affected by the presence of the
scalar field only for higher orders of the
slow-roll approximation, which are not included here.

\vskip 2mm The tensor modes can be decomposed into two independent polarizations,
\begin{eqnarray}
\label{eq:092}
\gamma_{\mathbf{k}ij}=\sum_{P=+,\times}^{ }e^P_{\mathbf{k}ij}\gamma^P_{\mathbf{k}},
\end{eqnarray}
where the polarization tensor $e^P_{ij}$
must satisfy the traceless and transversal conditions $e^P_{ii}=0$ and $k_ie^P_{ij}=0$,
and as the normalization condition we use $e^P_{ij}e^{P'*}_{ij}=\delta_{PP'}$.
The quantized tensor modes can be written in the form
\begin{eqnarray}
\label{eq:093}
\gamma_{\mathbf{k}ij}=\sum_{P=+,\times}^{ }
\left(e^P_{\mathbf{k}ij}\gamma^{(\textrm{cl})}_{\mathbf{k}}a^P_{\mathbf{k}}
+e^{P*}_{-\mathbf{k}ij}\gamma^{(\textrm{cl})*}_{-\mathbf{k}}a^{P\dag}_{-\mathbf{k}}\right),
\end{eqnarray}
where the creation and annihilation operators obey the standard commutation relations
and $\gamma^{(\textrm{cl})}$ denotes the classical solution of equation of motion
given by the tensor quadratic action (\ref{eq:033}).

\vskip 2mm
The first equation in (\ref{eq:047}) implies that
if the parameter $\eta_L$ defined in (\ref{eq:048}) is small,
and the parameter $\eta_T$ defined in the same manner,
\begin{eqnarray}
\label{eq:094}
\eta_T=\frac{\dot{c}_T}{c_TH},
\end{eqnarray}
must be small as well,
unless $\epsilon-p$ is much smaller than $\epsilon$.
In such case smallness of $\eta_T$ is an independent assumption.
For calculation of tensor power spectrum and bispectrum
we need no additional assumptions. We only have to take into account
restrictions which are consequence of the slow-roll approximation, including $p<\epsilon$,
and relations (\ref{eq:042})-(\ref{eq:047b}).
The equation for tensor perturbations then can be written as
\begin{eqnarray}
\label{eq:095}
\gamma^{(\textrm{cl})\prime\prime}_{\mathbf{k}}
-2\frac{1+\epsilon_c}{\tau}\gamma^{(\textrm{cl})\prime}_{\mathbf{k}}
+\left(k^2+4\frac{\epsilon_c-p_c}{\tau^2}c_{T,c}^2\right)
\gamma^{(\textrm{cl})}_{\mathbf{k}}=0,
\end{eqnarray}
where only terms up to the first order of the slow-roll approximation
have been kept, and notation follows the previous sections.
By solving this equation and matching the canonically normalized tensor mode,
$\gamma^{(\textrm{can})}_{\mathbf{k}}=\frac{1}{\sqrt{2}}M_\textrm{Pl}a\gamma^{(\textrm{cl})}_{\mathbf{k}}$,
to the free wave function of the Minkowski space vacuum, we find
\begin{eqnarray}
\label{eq:096}
\gamma^{(\textrm{cl})}_{\mathbf{k}}=-\sqrt{\frac{\pi}{2}}\frac{H_c}{M_\textrm{Pl}}
(1-\epsilon_c)(-\tau_c)^{-\epsilon_c}e^{i\frac{\pi}{2}\epsilon^{(\gamma)}_c}
(-\tau)^{\frac{3}{2}+\epsilon_c}
H^{(1)}_{\frac{3}{2}+\epsilon^{(\gamma)}_c}(-k\tau),
\end{eqnarray}
where $\epsilon^{(\gamma)}=(1-4c_T^2/3)\epsilon+4c_T^2p/3$ and if $\epsilon-p\sim\epsilon$ it can be rewritten as $(1+c_L^2)p-c_L^2\epsilon$.

\vskip 2mm The tensor power spectrum $\mathcal{P}_\gamma(k)$ is defined by
\begin{eqnarray}
\label{eq:097}
\left<\gamma_{\mathbf{k}_1ij}\gamma_{\mathbf{k}_2ij}\right>
=\frac{\mathcal{P}_\gamma(k_1)}{2k_1^3}
(2\pi)^5\delta^{(3)}(\mathbf{k}_1+\mathbf{k}_2),
\end{eqnarray}
where the late time two-point tensor function is
\begin{eqnarray}
\label{eq:098}
\left<\gamma_{\mathbf{k}_1ij}\gamma_{\mathbf{k}_2kl}\right>
=(2\pi)^3\delta^{(3)}(\mathbf{k}_1+\mathbf{k}_2)
\sum\limits_{P=+,\times}^{ }e^P_{\mathbf{k}_1ij}e^{P*}_{\mathbf{k}_1kl}
\frac{H_c^2}{M_\textrm{Pl}^2}k_1^{-3}
\left(\frac{\tau}{\tau_c}\right)^{2\epsilon_c}
\left(-k_1\tau\right)^{-2\epsilon^{(\gamma)}_c}.
\end{eqnarray}
As a result, the tensor spectral tilt is small,
\begin{eqnarray}
\label{eq:099}
n_T-1=-2\epsilon^{(\gamma)}_c=\frac{8}{3}c_{T,c}^2(\epsilon_c-p_c)-\epsilon_c.
\end{eqnarray}
For $c_T^2=(3/4)\epsilon/(\epsilon-p)$ the tensor power spectrum is flat, for $c_T^2>(3/4)\epsilon/(\epsilon-p)$ it is blue shifted and for $c_T^2<(3/4)\epsilon/(\epsilon-p)$ it is redshifted.
Furthermore, the tensor-to-scalar ratio is
\begin{eqnarray}
\label{eq:100}
r=\frac{\mathcal{P}_\gamma}{\mathcal{P}_\zeta}
=\frac{4c_{L}^5\epsilon^2}{\epsilon+\left(c_{L}^5-1\right)p},
\end{eqnarray}
being of the order $r\sim c_L^5\epsilon\sim\epsilon^2$ if $\epsilon-p\sim\epsilon\sim p$
and $c_L^5\sim\epsilon$, and it is amplified to the order of $\epsilon$ if $\epsilon-p$
is of the order $\epsilon^2$.
This does not contradict the observational restrictions \cite{planck2,bicep}. Enhancement of this quantity in case when $\epsilon-p\sim\epsilon^2$ seems to be similar to enhancement of the scalar non-linearity parameter $f_{\textrm{NL}}$ found in the previous section. In this case the reason is the structure of relation (\ref{eq:061}), relating the perturbation $\zeta$ to the perturbations of solid matter and scalar field, according to which contribution of solid matter to $\zeta$ is proportional to $(\epsilon-p)^{-1/2}$.

\vskip 2mm The tensor three-point function can be computed in the
same way as the scalar one in the previous section.
In order to do so, we need the tensor cubic action,
\begin{eqnarray}
\label{eq:101}
S^{(3)}_\gamma=\int d^4xa^3\left[
\frac{1}{4}M_\textrm{Pl}^2a^{-2}\gamma_{ij}\gamma_{kl}
\left(\gamma_{ik,jl}-\frac{1}{2}\gamma_{kl,ij}\right)
+\frac{F_Y}{27}\gamma_{ij}\gamma_{ik}\gamma_{jk}
\right].
\end{eqnarray}
Keeping only the leading order terms in the slow-roll
approximation, we find the three-point function in the form
\begin{eqnarray}
\label{eq:102}
& &\left<\gamma_{\mathbf{k}_1i_1j_1}(0)\gamma_{\mathbf{k}_2i_2j_2}(0)\gamma_{\mathbf{k}_3i_3j_3}(0)\right>
=\frac{16H_c^2(2\pi)^3\delta^{(3)}(\mathbf{k}_1+\mathbf{k}_2+\mathbf{k}_3)}{M_\textrm{Pl}^6\left(k_1k_2k_3\right)^3}
\cdot\\
& &\phantom{medzera}
\cdot\Bigg\{
\Bigg[-\frac{1}{4}M_\textrm{Pl}^2H_c^2\Pi_{i_1j_1ab}(\mathbf{k}_1)\Pi_{i_2j_2cd}(\mathbf{k}_2)
\Bigg(\Pi_{i_3j_3ac}(\mathbf{k}_3)(k_3)_b(k_3)_d-\nonumber\\
& &\phantom{medzera}
-\frac{1}{2}\Pi_{i_3j_3cd}(\mathbf{k}_3)(k_3)_a(k_3)_b\Bigg)
+5\textrm{ permutations}\Bigg]\Gamma_{k_1,k_2,k_3}+\nonumber\\
& &\phantom{medzera}
+\frac{2}{9}F_Y\Pi_{i_1j_1ab}(\mathbf{k}_1)\Pi_{i_2j_2ac}(\mathbf{k}_2)\Pi_{i_3j_3bc}(\mathbf{k}_3)\Xi_{k_1,k_2,k_3}\Bigg\},\nonumber
\end{eqnarray}
where $\Pi_{abcd}(\mathbf{k})=\sum_Pe^P_{ab\mathbf{k}}e^{P*}_{cd\mathbf{k}}$,
and $\Gamma_{k_1,k_2,k_3}$ and $\Xi_{k_1,k_2,k_3}$ are given by the integrals
\begin{eqnarray}
\label{eq:103}
\Gamma_{k_1,k_2,k_3}=\textrm{Re}\left\{\int\limits_0^\infty(i+k_1z)(i+k_2z)(i+k_3z)e^{i(k_2+k_2+k_3)z}z^{-2}dz\right\},
\end{eqnarray}
\begin{eqnarray}
\label{eq:104}
\Xi_{k_1,k_2,k_3}=\textrm{Re}\left\{\int\limits_0^\infty(i+k_1z)(i+k_2z)(i+k_3z)e^{i(k_2+k_2+k_3)z}z^{-4}dz\right\},
\end{eqnarray}
The first integral can be computed with the tilted integration contour,
as the integral (\ref{eq:082}), and it is of the form
\begin{eqnarray}
\label{eq:105}
\Gamma_{k_1,k_2,k_3}=\sum_ik_i-\frac{1}{2}\frac{\sum_{i\neq j}k_ik_j}{\sum_ik_i}
-\frac{\prod_ik_i}{\left(\sum_ik_i\right)^2},
\end{eqnarray}
but in the second integral a logarithmic divergence occurs
due to the lower limit of the integration interval,
similarly as in the integral (\ref{eq:083}).
If we replace the integration limits $(0,\infty)$
by $(-\tau_e,\infty)$, we find that the second integral
in the limit of small $\tau_e$ is
\begin{eqnarray}
\label{eq:106}
\Xi_{k_1,k_2,k_3}&=&\frac{1}{6}\left(\sum_ik_i\right)\left(\sum_{i\neq j}k_ik_j\right)
-\frac{4}{3}\prod_ik_i+\\
&+&\left[\frac{4}{9}-\frac{1}{3}\gamma_\textrm{EM}-\frac{1}{3}
\ln{\left(-\tau_e\sum_ik_i\right)}
\right]\sum_ik_i^3.\nonumber
\end{eqnarray}
It is dominated by
\begin{eqnarray}
\label{eq:106_1}
\Xi_{k_1,k_2,k_3}=\frac{1}{3}\mathcal{N}_\gamma\sum_ik_i^3,
\end{eqnarray}
where $\mathcal{N}_\gamma$ is a number of the order of number of e-folds,
and the results presented in what follows
were computed by using this relation instead of (\ref{eq:106}).

\begin{figure}[htb]
\centering
\includegraphics[scale=0.15]{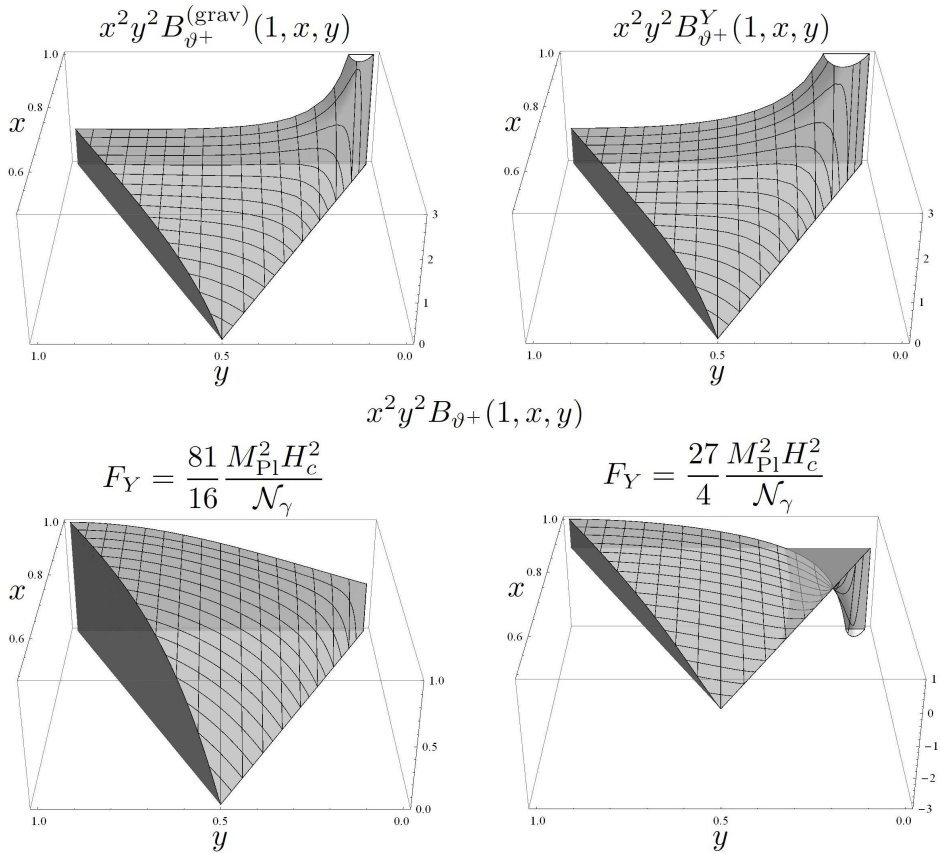}
\caption{
Shapes of the tensor bispectrum.
}
\label{fig:06}
\end{figure}

\vskip 2mm Conventionally one computes the three-point function
for polarization modes $\vartheta^P_{\mathbf{k}}$ defined as
\begin{eqnarray}
\label{eq:108}
\vartheta^P_{\mathbf{k}}=\gamma_{\mathbf{k}ij}e^{P*}_{ij}.
\end{eqnarray}
Using the properties of the polarization tensor,
the tensor three-point function (\ref{eq:102})
can be rewritten for the polarization
mode $\vartheta^+_{\mathbf{k}}$ as
\begin{eqnarray}
\label{eq:109}
\left<\vartheta^+_{\mathbf{k}_1}\vartheta^+_{\mathbf{k}_2}\vartheta^+_{\mathbf{k}_3}\right>
&=&-\frac{16H_c^2}{M_\textrm{Pl}^2}(2\pi)^3\delta^{(3)}(\mathbf{k}_1+\mathbf{k}_2+\mathbf{k}_3)\cdot\\
& &\cdot\left(
\frac{1}{4}M_\textrm{Pl}^2H_c^2\mathcal{G}^{+++}_{k_1,k_2,k_3}\Gamma_{k_1,k_2,k_3}
-\frac{2}{9}F_Y\mathcal{X}^{+++}_{k_1,k_2,k_3}\Xi_{k_1,k_2,k_3}
\right),\nonumber
\end{eqnarray}
where
\begin{eqnarray}
\label{eq:110}
\mathcal{G}^{+++}_{k_1,k_2,k_3}&=&\frac{\left(\sum_ik_i\right)^2}{2}\mathcal{X}^{+++}_{k_1,k_2,k_3},
\\
\mathcal{X}^{+++}_{k_1,k_2,k_3}&=&\frac{\left(\sum_ik_i\right)^3}{64\left(\prod_ik_i\right)^5}
\left[\left(\sum_ik_i\right)^3-2\left(\sum_ik_i\right)\left(\sum_{i\neq j}k_ik_j\right)+8\prod_ik_i\right].\nonumber
\end{eqnarray}

\vskip 2mm The tensor bispectrum $B_{\vartheta^+}(k_1,k_2,k_3)$ consists of two parts,
\begin{eqnarray}
\label{eq:111}
\left<\vartheta^+_{\mathbf{k}_1}\vartheta^+_{\mathbf{k}_2}\vartheta^+_{\mathbf{k}_3}\right>
&=&(2\pi)^3\delta^{(3)}(\mathbf{k}_1+\mathbf{k}_2+\mathbf{k}_3)B_{\vartheta^+}(k_1,k_2,k_3)=\\
&=&(2\pi)^3\delta^{(3)}(\mathbf{k}_1+\mathbf{k}_2+\mathbf{k}_3)
\left[B^{(\textrm{grav})}_{\vartheta^+}(k_1,k_2,k_3)+\mathcal{N}_\gamma F_YB^{Y}_{\vartheta^+}(k_1,k_2,k_3)\right].
\nonumber
\end{eqnarray}
The behaviour of both parts as well as of their sum for two values of $F_Y$
is depicted in fig. \ref{fig:06}.
The first part of the bispectrum $B^{(\textrm{grav})}_{\vartheta^+}(k_1,k_2,k_3)$,
given by the non-linear structure of the scalar curvature in general relativity,
can be found in most inflationary models,
in particular single-field ones, see \cite{gao}.
The second part $B^{Y}_{\vartheta^+}(k_1,k_2,k_3)$
represents the effect of the solid,
while the presence of the scalar field in our model
affects the tensor bispectrum only
in higher orders of slow-roll approximation,
which are not included in our work.

\vskip 2mm Both parts of the tensor bispectrum peak
in the squeezed limit.
The overall bispectrum does not peak in this limit
if $F_Y=F_Y^{(\textrm{lim})}=(81/16)M_\textrm{Pl}^2H_c^2\mathcal{N}_\gamma^{-1}$,
and it peaks in the equilateral limit instead.
For $F_Y<F_Y^{(\textrm{lim})}$ the peak in the squeezed limit
has the same sign as the bispectrum
in the equilateral limit and for
$F_Y>F_Y^{(\textrm{lim})}$ their relative sign is minus.
This is demonstrated in the third and fourth panel
of the fig. \ref{fig:06}.
The tensor bispectrum is always zero in the folded limit.

\section{Conclusion}

In this paper we have studied a model in which inflation is driven not only by a solid as in \cite{endlich}, but also by a scalar field $\varphi$ with standard kinetic term. The object defining the model is the potential $F(\varphi,X,Y,Z)$, where the quantities $X$, $Y$ and $Z$ defined in (\ref{eq:004}) describe the solid. The model represents the most straightforward combination of solid inflation and the basic single-field inflationary model. It can be considered as, for instance, a simple toy model of interactions of fields driving the solid inflation with fields of an effective field theory of the standard model.

\vskip 2mm Due to the additional degree of freedom, the slow-roll parameter $\epsilon$ is a function of two independent parameters $p$ and $q$ defined in (\ref{eq:011}), which, in principle, allows for a wide range of inflationary scenarios. However, we have restricted ourselves to the special case such that both $p$ and $q$ are small, being of the same order as the slow-roll parameter. As a consequence, the scalar field mass squared $-F_{\varphi\varphi}$ is of the first order in the slow-roll parameter, which leads to the relation (\ref{eq:047}) between the transversal sound speed $c_T$ and the longitudinal sound speed $c_L$. Moreover, in case that $F_{X\varphi}$ has a special form given in (\ref{eq:fico}) the analysis of the cosmological perturbations can be treated analytically, since equations of motion for two scalar perturbations present in our model become decoupled if $\rho$ is replaced by $\mathcal{U}$.

\vskip 2mm Under assumptions adopted above the scalar spectrum is nearly flat and for the scalar bi\-spectrum different shapes are allowed. The reason is that there are three independent parameters of the theory which are not necessarily suppressed by the slow-roll parameters, $F_Y$, $\tilde{F}$ defined in (\ref{eq:076}) and the longitudinal sound speed. We computed the scalar bispectrum only in the leading order of the slow-roll approximation which does not require $F_{X\varphi}$ to be of the special form given by (\ref{eq:fico}). In solid inflation the bispectrum peaks in the squeezed limit with an anisotropic dependence on how the limit is approached. This applies also for our combined model, unless $\tilde{F}/F_Y=(5/6)\mathcal{N}_\zeta^{-1}c_L^{-2}$, when the bispectrum peaks in the equilateral limit instead. The non-linearity parameter is of the order $f_\textrm{NL}\sim (F_Y/F)c_L^{-2}\epsilon^{-1}$, the same as for the solid inflation without scalar field, or $f_\textrm{NL}\sim\mathcal{N}_\zeta(\tilde{F}/F)\epsilon^{-1}$, and is amplified by a factor of the order $\epsilon^{-2}$ when $\epsilon-p$ is of the order $\epsilon^2$, i.e. when the contribution of the solid matter to the overall stress-energy tensor is much smaller than the contribution from the scalar field. In this case the relation (\ref{eq:047}) between the sound speeds $c_T$ and $c_L$ is not valid. The case when $\epsilon-p$ is of order $\epsilon^{5/2}$ or smaller is excluded, since the sound speeds would be superluminal or imaginary.

\vskip 2mm The tensor power spectrum is nearly flat with spectral tilt given by the slow-roll parameters $\epsilon$ and $p$ and the longitudinal sound speed. The tensor-to-scalar ratio is of the order $r\sim c_L^5\epsilon\sim\epsilon^2$ if $\epsilon-p\sim\epsilon\sim p$, while for $\epsilon-p$ being of the order $\epsilon^2$ we have $r\sim\epsilon$, which is in agreement with the observational restrictions. In our model the tensor bispectrum computed in the leading order of the slow-roll approximation does not differ from the tensor bispectrum in solid inflation. It is affected by presence of the scalar field only in the higher orders of the slow-roll approximation, which are not included in our work.

\vskip 2mm Although the part of the analysis of cosmological perturbations presented in our work concerning scalar power spectrum tilt is valid only when parameters of the theory are fine tuned so that (\ref{eq:fico}) is satisfied, the problem can be studied also without this restriction. Of course, this would demand more complicated or even nonanalytic treatment. We leave this for future work.

\vskip 1cm
\noindent {\bf{\Large Acknowledgments}}
\vskip 5mm \noindent I would like to thank Vladim\'ir Balek for useful discussions and valuable comments, and also for careful reading of the manuscript. The work was supported by the grants VEGA 1/0985/16 and UK/36/2017.

\end{document}